\documentclass[11pt]{article}

\usepackage {hyperref}
\usepackage {color}
\usepackage[pdftex]{graphicx}
\usepackage{amsmath}
\usepackage{amsfonts}
\usepackage{amssymb}
\usepackage{epstopdf}
\usepackage{color}
\usepackage{amsmath,amssymb}
\usepackage{slashed}
\usepackage{xcolor} 
\usepackage{graphicx}
\usepackage{url}
\usepackage{multirow}
\usepackage[noadjust]{cite}
\usepackage{filecontents}
\usepackage{lineno}
\usepackage{subfig}
\usepackage{parskip}
\usepackage[italic]{hepnames}

\usepackage{perpage} 
\MakePerPage{footnote} 

\newcommand{\be}{\begin{equation}}
\newcommand{\ee}{\end{equation}}
\newcommand{\bea}{\begin{eqnarray}}
\newcommand{\eea}{\end{eqnarray}}
\def\Re{{\cal R \mskip-4mu \lower.1ex \hbox{\it e}\,}}
\def\Im{{\cal I \mskip-5mu \lower.1ex \hbox{\it m}\,}}

\def\tev{\,{\ifmmode\mathrm {TeV}\else TeV\fi}}
\def\gev{\,{\ifmmode\mathrm {GeV}\else GeV\fi}}
\def\mev{\,{\ifmmode\mathrm {MeV}\else MeV\fi}}
\def\to{\rightarrow}
\usepackage{lipsum}
\renewenvironment{abstract}{%
\hfill\begin{minipage}{0.95\textwidth}
\rule{\textwidth}{1pt}}
{\par\noindent\rule{\textwidth}{1pt}\end{minipage}}

\begin{document}

\begin{center}
\vspace*{10mm}

\vspace{1cm}
{\huge \bf  Photon initiated single top quark production via flavor-changing neutral currents at the LHC  } \\
\vspace{1cm}

{\large \bf Reza Goldouzian\footnote{email: Reza.Goldouzian@cern.ch} and Barbara Clerbaux \footnote{email: Barbara.Clerbaux@cern.ch }   }

 \vspace*{.3cm}
   {\emph{Interuniversity Institute for High Energies (IIHE),\\
Physique des particules \'el\'ementaires,\\
Universit\'e Libre de Bruxelles, ULB, 1050, Brussels, Belgium} }
   
   {\small (Dated: September, 2016) }

\end{center}

\vspace*{10mm}
\begin{abstract}
Single top quark production is a powerful process to search for new physics signs.
In this work we propose and investigate a search for top quark flavor changing neutral currents (FCNC) via a photon using direct single top quark production events in proton-proton collisions at the LHC at CERN. We show that the direct single top quark final state can provide constraints on the strengths of tq$\gamma$ (top-quark-$\gamma$) and tqg (top-quark-gluon) FCNC couplings simultaneously. 
Results of a search for direct single top quark production at the LHC at a center-of-mass energy of 8 TeV performed by the ATLAS collaboration  are used to set first experimental limits on the anomalous FCNC top decay branching fractions ${\cal B}$(t$\rightarrow$ u$\gamma$) $<$ 0.05\% and ${\cal B}$(t$\rightarrow$ c$\gamma$) $<$ 0.14\% via direct single top quark production. Finally, the sensitivity  of the proposed channel for probing the tq$\gamma$ couplings at 13 TeV is presented.
\end{abstract}

\vspace*{10mm}
\clearpage

\section{Introduction}\label{Introduction}
The top quark, with a mass of 173.3 ($\pm$0.8) GeV \cite{ATLAS:2014wva} close  to  the  electroweak  symmetry  breaking scale, is considered as an excellent probe to search for new physics beyond the standard  model (BSM). 
Contributions of new particles or interactions of BSM physics  can modify the production rate and properties of the top quark expected in the standard model (SM). 
In the SM, flavour changing neutral currents (FCNC) are forbidden at tree level and are suppressed  at  higher orders  due  to  the Glashow-Iliopoulos-Maiani (GIM) mechanism \cite{Glashow:1970gm}.
The branching fractions of the top quark FCNC decays to up-type quarks and a gluon, photon, Z or Higgs boson predicted in the SM are extremely small (of the order of $10^{-17}$ to $10^{-12}$) and far beyond the current sensitivity of the experiments. 
On the other hand, these tiny branching fractions can be enhanced significantly in a number of new physics models for certain regions of the parameter space and could lie at the edge of present-day experimental limits (of the order of $10^{-4}$ to $10^{-6}$) \cite{AguilarSaavedra:2004wm}.
\\
\\
Enhanced top quark FCNC interactions can be described in a model-independent way using the effective Lagrangian approach. The most general effective Lagrangian for top quark FCNC interactions derived from dimension-six operators can be written as \cite{AguilarSaavedra:2004wm}:
\begin{eqnarray}
-\mathcal{L}_{eff} &=& e \frac{\kappa_{\text{q}\gamma}}{\Lambda} \bar{q} i\sigma^{\mu\nu}k_{\nu} [\gamma_{L} P_{L}+\gamma_{R} P_{R}]tA_{\mu} + g_{s} \frac{\kappa_{\text{qg}}}{\Lambda}\bar{q} i\sigma^{\mu\nu}k_{\nu} [g_{L} P_{L}+g_{R} P_{R}]T^a tG_{a\mu}  \nonumber\\
&+&\frac{g}{2 \text{cos}{\theta_{w}}} \frac{\kappa_{\text{qZ}}}{\Lambda} \bar{q}i\sigma^{\mu\nu}k_{\nu} [z_{L} P_{L}+z_{R} P_{R}]tZ_{\mu}  
+\kappa_{\text{qH}}\bar{q} [h_{L} P_{L}+h_{R} P_{R}]tH + h.c.
\label{lagrangy}
\end{eqnarray}
Where $\kappa_{\text{q}\gamma}$, $\kappa_{\text{qg}}$, $\kappa_{\text{qZ}}$ and $\kappa_{\text{qH}}$ are the real and positive parameters which determine the  strength of the new top quark FCNC interactions with a photon, gluon, Z and Higgs boson, respectively.  In equation \ref{lagrangy}, $e$ is the electron electric charge, $g$ and  $g_{s}$ are the weak and strong coupling constants, $\theta_{w}$ is the Weinberg angle, $P_{L (R)}$ denotes the
left(right)-handed projection operator, $\sigma^{\mu\nu} = \frac{1}{2}[\gamma^{\mu},\gamma^{\nu}]$, $q$ represents the  spinor fields of up and charm quark, $t$ is the top quark spinor field, $k$ is the momentum of the intermediate gauge bosons or Higgs. The new FCNC interactions can couple to the left and right components of the quark fields differently which are parametrized by $\gamma_{L,R}$, $g_{L,R}$, $z_{L,R}$ and $h_{L,R}$ and are normalized as  $|X_{L}|^2 + |X_{R}|^2=1$. The new physics scale $\Lambda$  has a dimension of energy and is conventionally taken as the top quark mass. This effective Lagrangian is valid below the new physics scale and breaks down above this energy scale where the new physics is expected to appear.

\begin{figure}[th]
\centering 
\subfloat[][]{
\includegraphics[width=.35\textwidth]{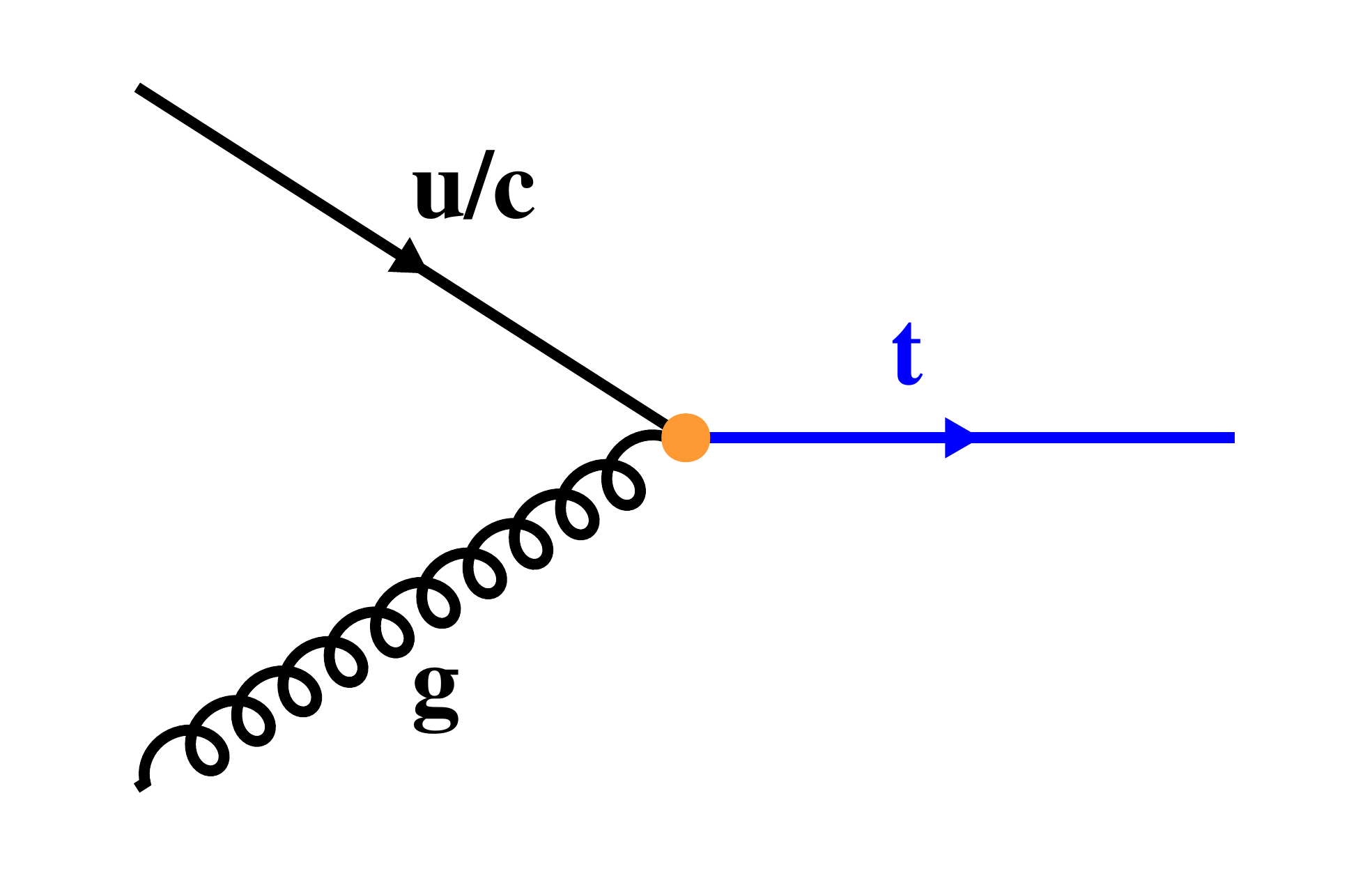}{}}
\subfloat[][]{
\includegraphics[width=.35\textwidth]{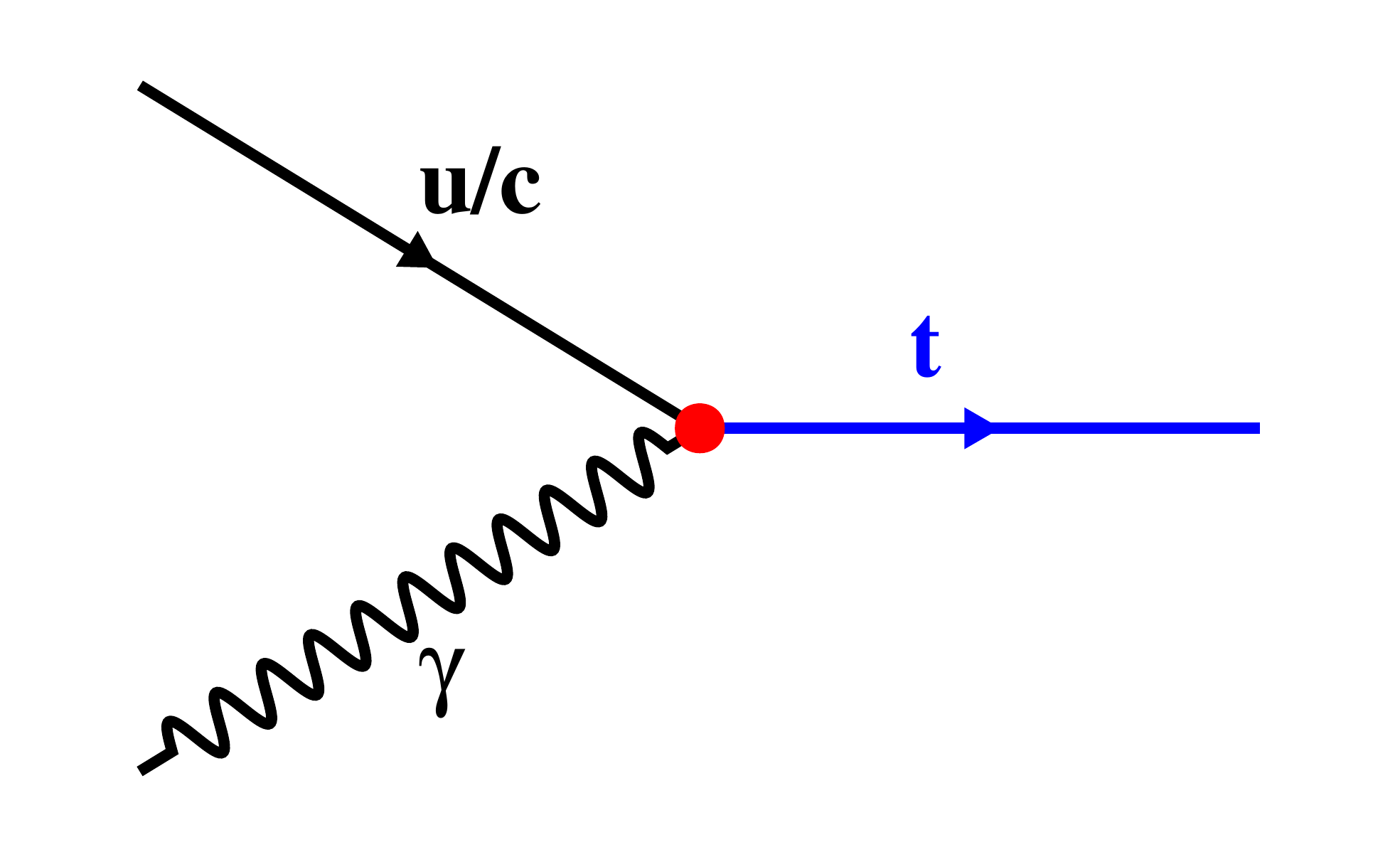}{}} \\
\subfloat[][]{
\includegraphics[width=.23\textwidth]{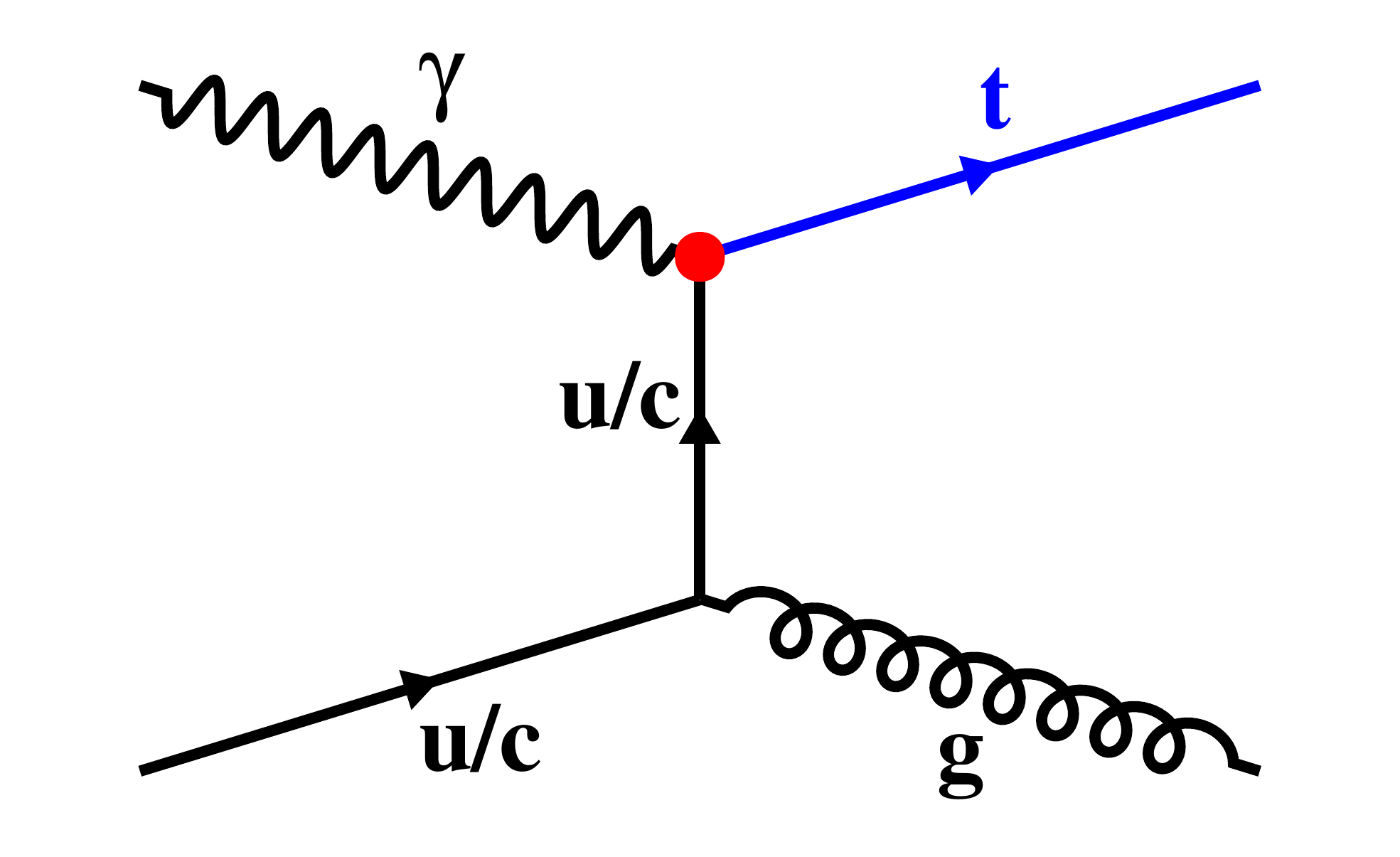}{}}
\subfloat[][]{
\includegraphics[width=.23\textwidth]{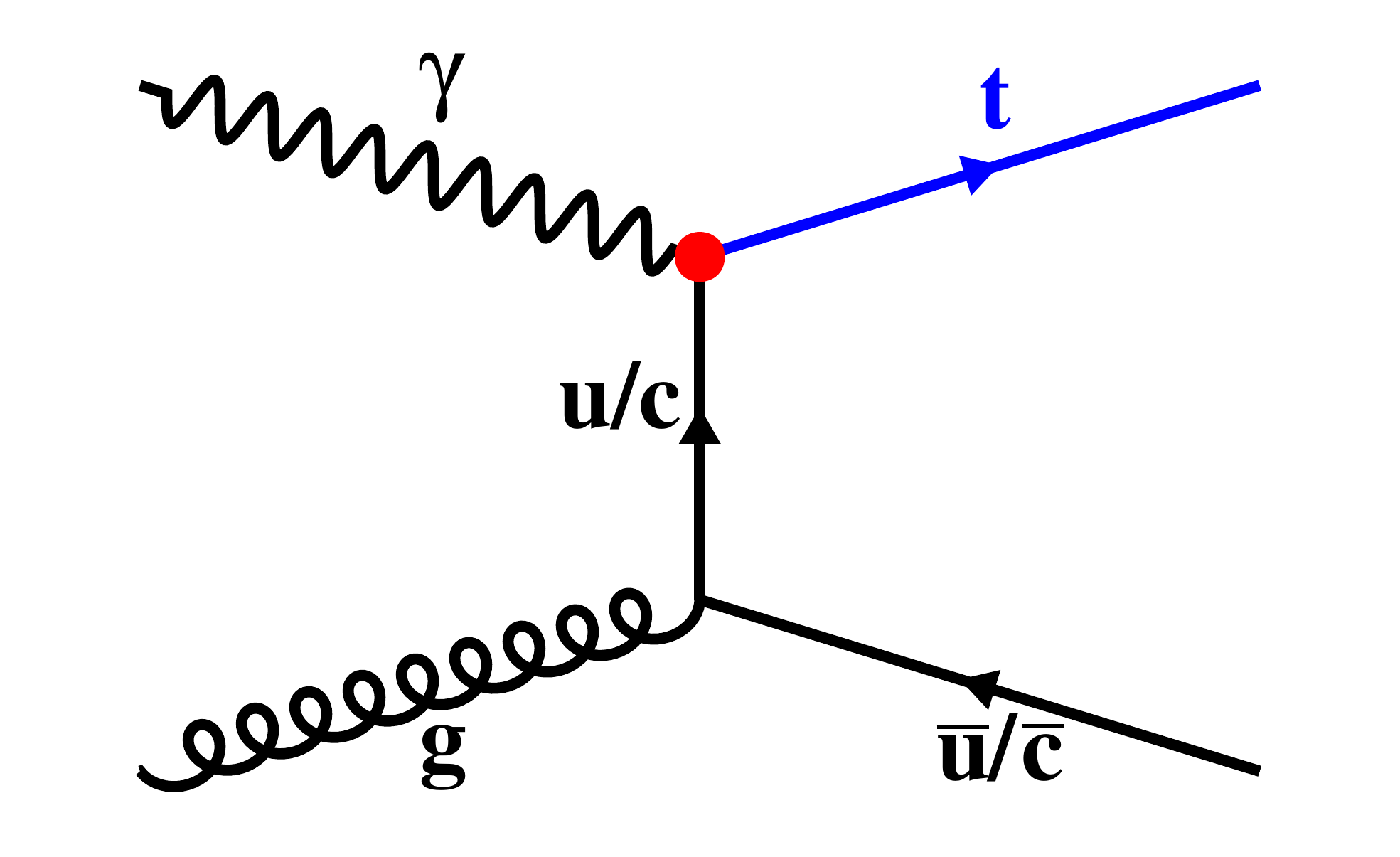}{}}
\subfloat[][]{
\includegraphics[width=.23\textwidth]{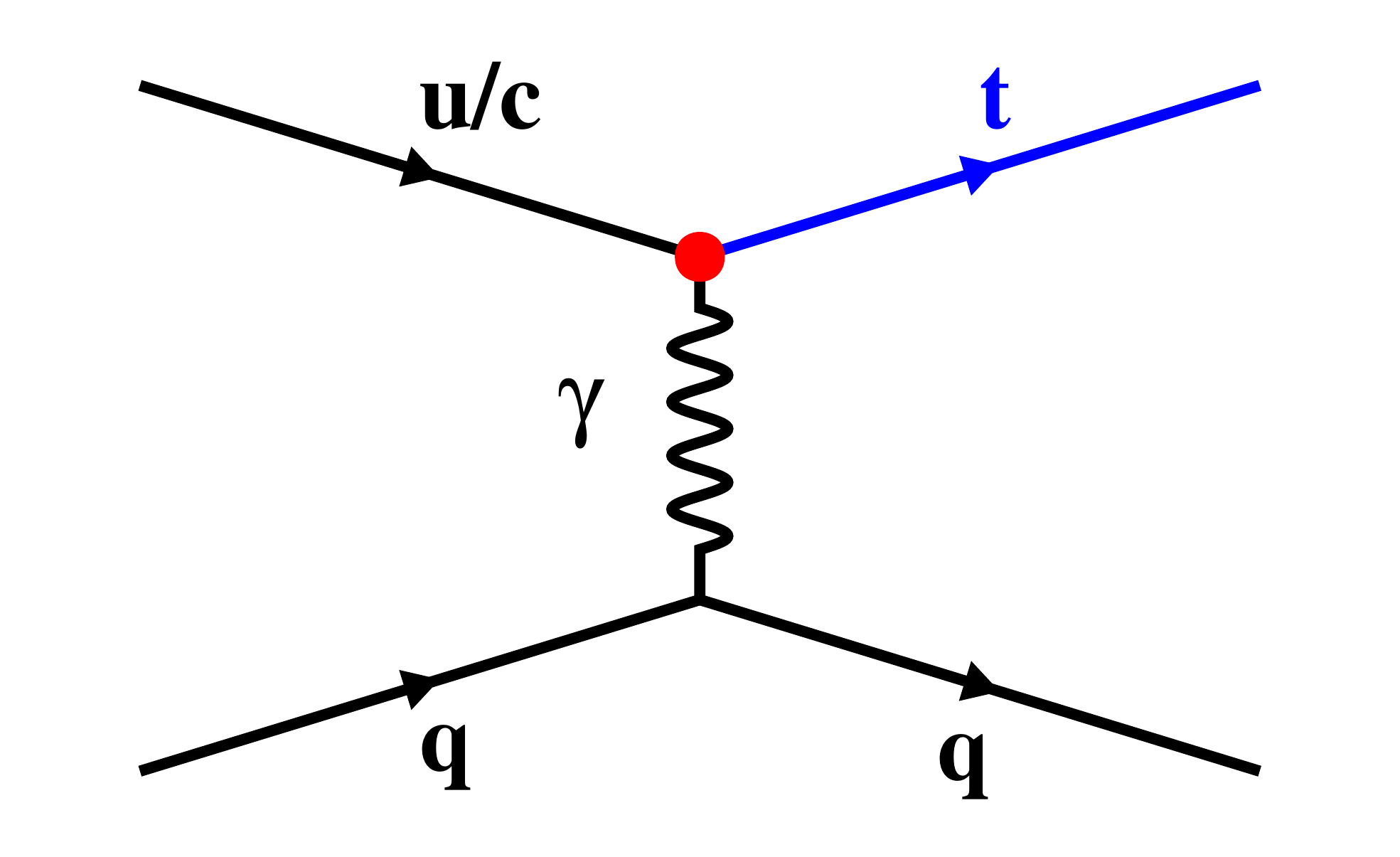}{}}
\subfloat[][]{
\includegraphics[width=.23\textwidth]{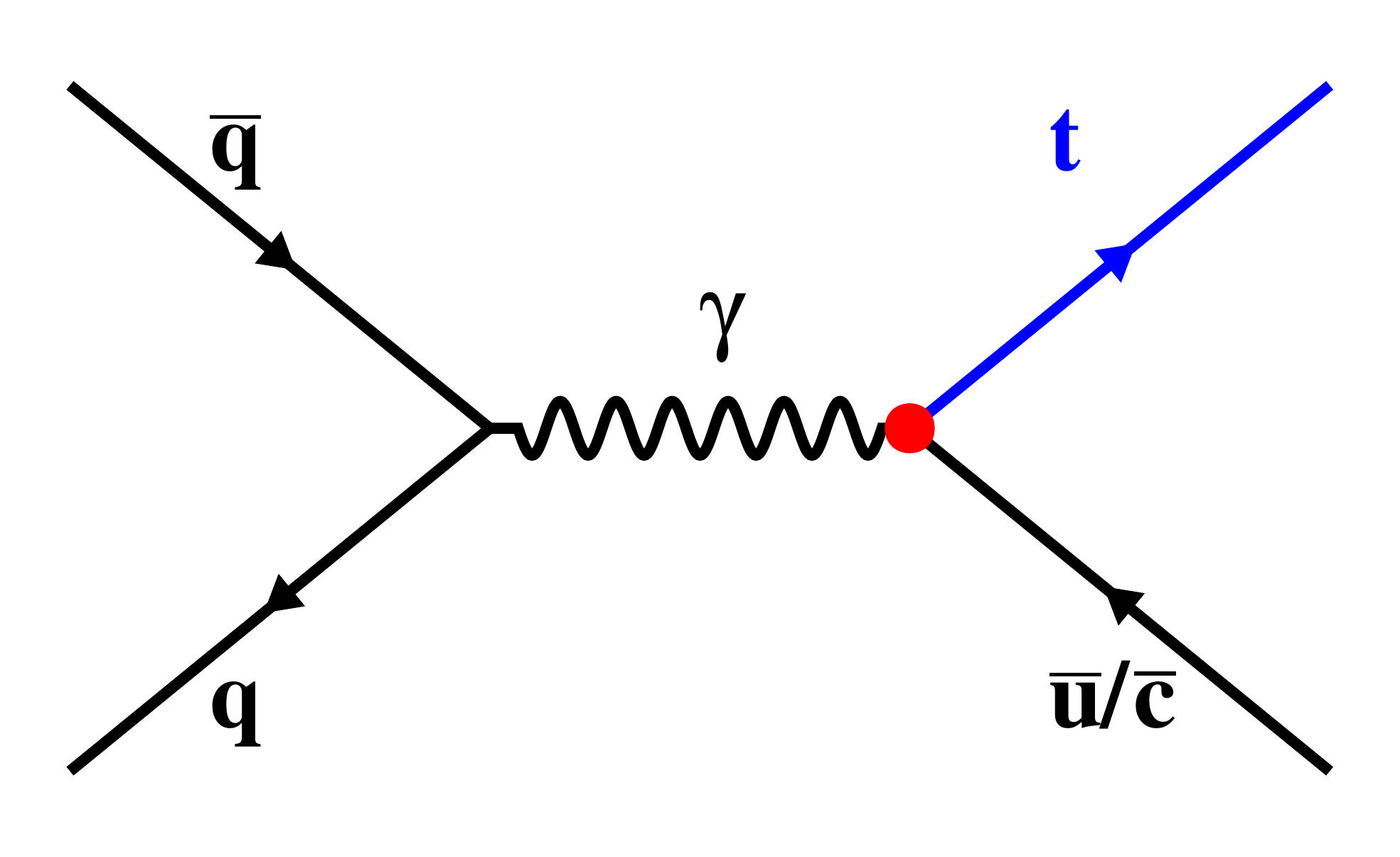}{}}  \\
\subfloat[][]{
\includegraphics[width=.23\textwidth]{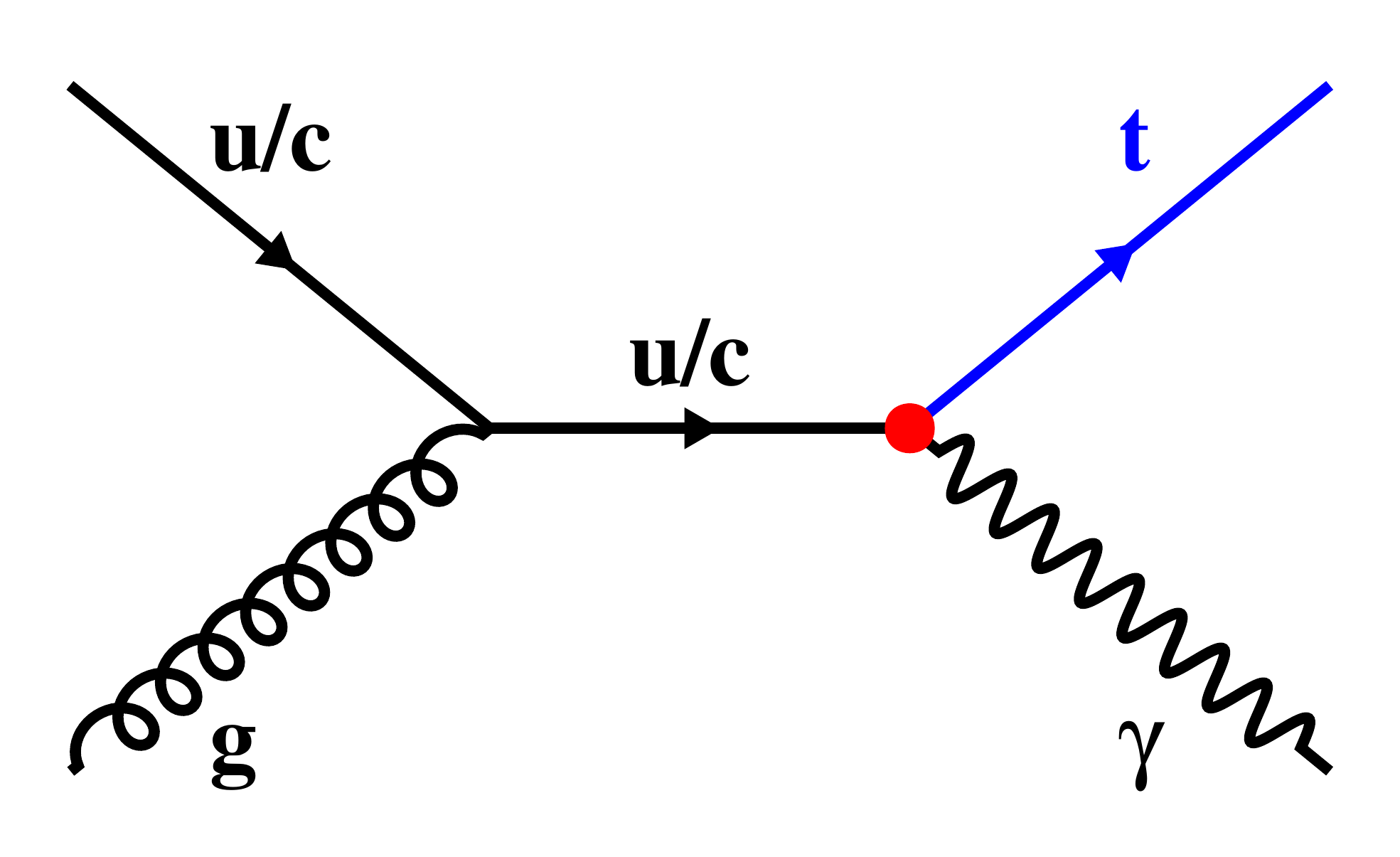}{}}
\subfloat[][]{
\includegraphics[width=.23\textwidth]{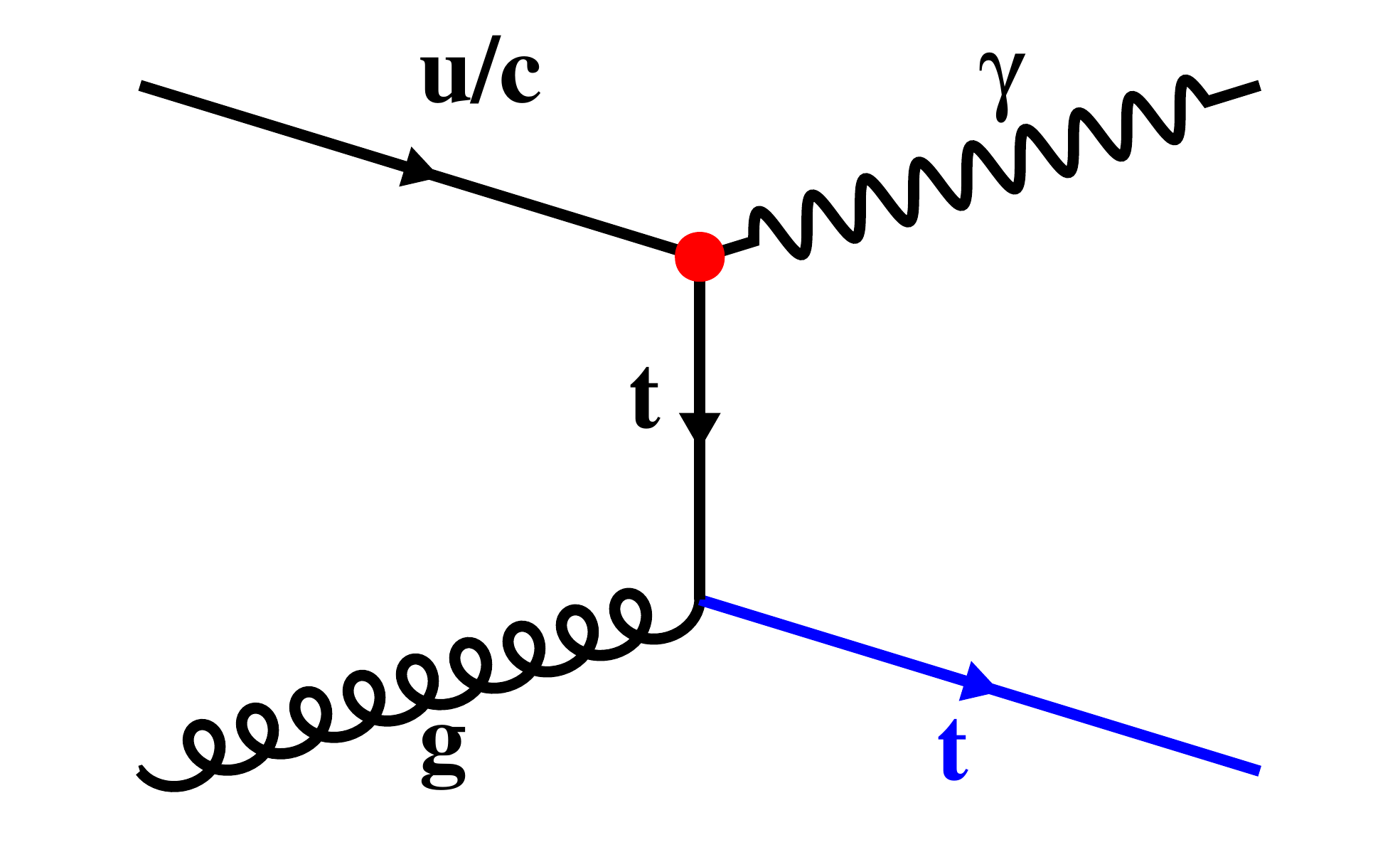}{}}
\subfloat[][]{
\includegraphics[width=.23\textwidth]{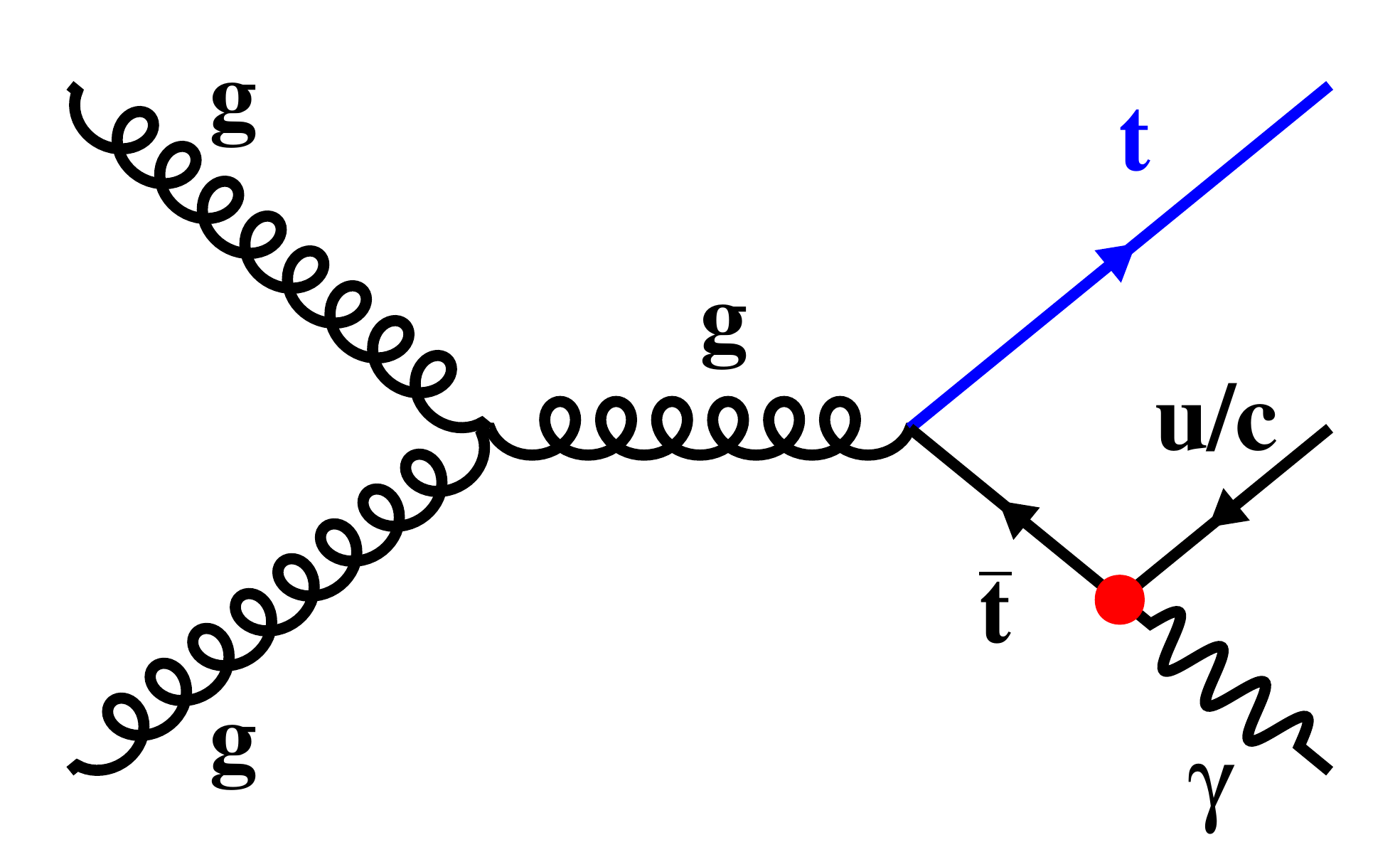}{}}
\hfill
\caption{Representative leading order Feynman diagrams describing the direct single top quark production (a-b), single top quark production in association with a jet (c-f), single top quark production in association with a photon (g-h) and  top quark decay to a photon and up-type quark from the SM t$\bar{\text{t}}$ production (i), via FCNC interactions in proton-proton collisions.}
\label{feyn}
\end{figure}

Many final states with various sensitivities are proposed by phenomenologists to search for FCNC effects involving top quark \cite{AguilarSaavedra:2004wm}. The FCNC processes can manifest themselves in decays of the top quark: t $\rightarrow$ qX with q = u, c and X = $\gamma$, g, Z, H. These searches benefit from the large number of top quark events produced in high energy hadron colliders. In addition to the anomalous top quark decays, the top quark FCNC interactions can lead to the anomalous production of single top quark or other final states. For example, tq$\gamma$ anomalous interaction can be probed in single top and photon \cite{AguilarSaavedra:2004wm}, same-sign double top \cite{Goldouzian:2014nha} or di-photon \cite{Khatibi:2015aal} production in proton-proton collisions.
\\
\\
Searches for FCNC top quark couplings have been pursued by various experiments for many years. From LEP data, only weak constraints on the tq$\gamma$ and tqZ couplings are reported through the search for anomalous single top quark production  \cite{Achard:2002vv}. Using HERA data, the limits on the  tq$\gamma$ FCNC couplings was improved, leading to a sensitivity comparable to the one of tqZ coupling \cite{Abramowicz:2011tv}. Given the large number of top quark pairs produced at the Tevatron, FCNC top quark  decay channels were used to probe the tq$\gamma$ and the tqZ  couplings \cite{Abe:1997fz,Abazov:2011qf}. Among the top quark FCNC decay channels, the t $\rightarrow$ qg channel suffers from a large QCD multijet background in hadron colliders. Therefore, direct single top quark signatures were chosen to search for tqg FCNC couplings \cite{Aaltonen:2008qr}. 
In figure 1 (a), the Feynman diagram for the direct single top quark production via anomalous tqg interaction is shown. Another channel used by Tevatron to search for tqg FCNC couplings is single top quark production in association with a jet. This final state is  similar to the SM single top quark production and is less sensitive compared to the direct single top production \cite{Abazov:2010qk}.  
The discovery of the SM-like H scalar boson at the LHC made possible the search for FCNC top quark couplings associated with a H boson \cite{Aad:2012tfa,Chatrchyan:2012xdj}. The ATLAS and CMS experiments have performed searches for FCNC top quark decays involving the H boson in various H decay modes \cite{Goldouzian:2014xfa}. 
In addition to new constraints on the anomalous tqH couplings, the ATLAS and CMS experiments have significantly improved previous exclusion limits on the other FCNC couplings with Run I data. The best limits on the branching fraction of the FCNC top quark decays obtained at the LHC are summarized in table \ref{FCNCexpresults}.

\begin{table}[ht]
\centering
\begin{tabular}{lll}
\hline
 & q = u     & q = c     \\
\hline
\hline
${\cal B}$(t$\rightarrow$ qg) & 0.004\% \cite{Aad:2015gea} & 0.020\% \cite{Aad:2015gea}\\
${\cal B}$(t$\rightarrow$ q$\gamma$) & 0.013\% \cite{Khachatryan:2015att} & 0.17\% \ \ \cite{Khachatryan:2015att} \\
${\cal B}$(t$\rightarrow$ qZ) & 0.05\%  \ \ \cite{Chatrchyan:2013nwa}      &   0.05\%  \ \  \cite{Chatrchyan:2013nwa}   \\
${\cal B}$(t$\rightarrow$ qH) &  0.42\% \ \   \cite{CMS:2015xqa}    &  0.46\% \ \ \cite{Aad:2015pja}   \\  
\hline
\end{tabular}
\caption{The most stringent observed upper limits at 95\% confidence level on the branching fraction of FCNC top quark  decays obtained at the LHC. }
\label{FCNCexpresults}
\end{table}

The LHC is also known to be a $\gamma \gamma$ and $\gamma$p collider. This feature offers a unique possibility to test the electroweak properties of the SM and to probe some BSM physics scenarios \cite{deFavereaudeJeneret:2009db}. The  photon emitted by the proton can interact with an up-type quark in the other proton via anomalous tq$\gamma$ interaction and leads to direct single top quark production. The leading order Feynman diagram of direct single top quark photoproduction is shown in figure 1 (b). Direct single top quark photoproduction can be separated in two classes of events: the elastic photon emission, i.e.  the proton emits a photon and remains intact; and the inelastic photon emission, i.e. the proton dissociates after emitting the photon. 
For elastic photoproduction, one proton is scattered with some energy loss and could be detected in spectrometers placed in the very forward region close to the beam line. By tagging the deflected proton in very forward spectrometers, photon initiated events can be clearly distinguished from the large background of inelastic proton-proton interactions. In addition, due to the presence of a color flow in inelastic pp interactions, the energy measured in the forward region is higher than for elastic photon initiated interactions. These two experimental signatures can be used to reduce the partonic background and to extract the photon initiated signal. However, the presence of extra pp interactions per bunch crossing (pile-up) in high luminosity pp collisions disables this method.
\\
\\
In the literature, photoproduction of the top quark is proposed in order to probe various properties of the top quark \cite{deFavereaudeJeneret:2009db,Fayazbakhsh:2015xba}. In refs. \cite{deFavereaudeJeneret:2008hf,Ovyn:2008gs,Sun:2014qoa,Koksal:2013fta,Inan:2014mua}, direct single top quark production channel through the elastic photon emission at LHC is studied to probe the tq$\gamma$ FCNC interactions.      
In the present paper, we propose and study the photoproduction of single top quark via both elastic and inelastic photon emissions for probing the anomalous FCNC tq$\gamma$ interactions at the LHC. The sensitivity of the proposed channel will be studied and first experimental limits will be set on the anomalous  tq$\gamma$  couplings using results obtained by the ATLAS experiment in a search for single top quark production via FCNC at 8 TeV \cite{Aad:2015gea}.
The organization of the paper is as follow. In section 2, the cross section calculation for photon initiated processes in pp collisions is discussed and numerical results for the single top quark photoproduction cross section are presented. In section 3, we perform an analysis similar to the one of ATLAS  and set experimental limits to the tq$\gamma$ couplings in section 4. Results on prospects at $\sqrt{s}$ = 13 TeV are presented in section 5. Finally conclusions are reported in section 6.

\section{Single top quark photoproduction cross section}

\subsection{Calculation setup}
As discussed in section \ref{Introduction}, the LHC offers the possibility to study $\gamma$q and $\gamma$g interactions, and single top quark can be produced in the presence of FCNC tq$\gamma$ couplings (see figure 1 (b)). In this process, both elastic and inelastic photon emissions contribute to the total cross section, with different signatures in the final state.
The $\gamma$p $\rightarrow$ t  cross section in which a quasi-real photon is emitted from the proton can be described by the equivalent photon approximation
(EPA) \cite{Budnev:1974de}. The EPA factorizes the cross section of the photon induced process in high energy pp collisions into the cross section of $\gamma$p folded with the photon flux. The photon flux, $N$($x$, $Q^2$), is a function of the virtuality $Q^2$  and the energy of the emitted photon (or its energy ratio to the hadron's energy, $x=\frac{E_\gamma}{E}$): 
\begin{eqnarray}
d\sigma_{\text{pp}} = \sigma_{\gamma \text{p}} (x,s) dN(x,Q^2). 
\label{dn}
\end{eqnarray}
If the photon is emitted from a nucleon which is not considered as a point-like particle, electromagnetic form factors should be taken into account in the photon flux calculation. Therefore,  the photon flux from a proton can be written as \cite{deFavereaudeJeneret:2008hf,Ovyn:2008gs,Sun:2014qoa,Koksal:2013fta,Inan:2014mua,Budnev:1974de}:
\begin{eqnarray}
dN(x,Q^2) = \frac{\alpha}{\pi} \frac{dQ^2}{Q^2} \frac{dx}{x} \left[ \left( 1-x \right) \left(1-\frac{Q^2_{min}}{Q^2}\right) F_E + \frac{x^2}{2} F_M \right], 
\label{EPA}
\end{eqnarray}
\begin{gather*}
F_E = \frac{4m_{\text{p}}^2 G_E^2 + Q^2 G_M^2}{4m_{\text{p}}^2 + Q^2}, \ \ F_M=G_M^2, \ \ Q^2_{min} \approx m_{\text{p}}^2 \frac{x^2}{1-x}, \\
G_E(Q^2) = \frac{ G_M(Q^2)}{\mu_{\text{p}}} = (1+\frac{Q^2}{0.71 {\text{GeV}}^2})^{-2},\\
\end{gather*}
where $m_{\text{p}}$ is the mass of the proton, $\alpha$ is fine-structure constant, $\mu_{\text{p}}$ is the proton magnetic moment,  $F_E$ and $F_M$ are the electric and magnetic form factors, respectively.
\\
\\
The above description also applies when the photon emission is inelastic by introducing structure functions instead of the electromagnetic form factors and redefining the minimum photon virtuality as:
\begin{gather*}
Q^2_{min} \approx \left[ m_X^2 \frac{x^2}{1-x} - m_{\text{p}}^2 \right] x, \\
\end{gather*}  
where $m_X$ is the mass of the system produced after dissociation of the proton. Therefore, the cross section calculation requires the inclusion of the photon distribution functions in the  proton, also called the photon PDF. The Photon PDF was introduced for the first time more than ten years ago by the MRST collaboration \cite{Martin:2004dh}.
The CTEQ and NNPDF collaborations have proposed a more general parametrization to describe the photon PDF and have used constraints coming from the HERA and LHC data, respectively \cite{Schmidt:2015zda,Ball:2013hta}. 
\\
\\
We implemented the effective Lagrangian defined in equation \ref{lagrangy} into the {\sc FeynRules} program \cite{Alloul:2013bka,Degrande:2011ua} and used the {\sc MadGraph\_}a{\sc mc@NLO} program for the cross section calculation and event generation \cite{Alwall:2014hca}. The EPA, implemented in {\sc MadGraph\_}a{\sc mc@NLO}, is used to calculate the cross section  of the direct single top quark production from  elastic photon emission. For the inelastic photon emission, three PDF sets have been considered: MRST\_QED, CT14\_QED and NNPDF23\_QED, which include a photon component.\footnote{The precise name of the 3 PDF sets in the official LHAPDF 6.1 PDF sets are 'MRST2004qed', 'CT14qed\_proton' and 'NNPDF23\_nlo\_as\_0119\_qed' \cite{Buckley:2014ana}.} The top quark mass is set to 173 GeV in all parts of this study.

\subsection{Signal channels}

Effects of the FCNC tq$\gamma$ coupling  could be significant in some interesting single top quark production processes in pp collisions \footnote{The contribution of anti-top quark is considered in all part of this study but is not mentioned in the text.} which are as follows:
\begin{itemize}
\item direct single top quark production (pp $\rightarrow$ t), the representative Feynman diagram is shown in figure 1 (b);
\item single top + jet production (pp $\rightarrow$ t+jet), representative Feynman diagrams are shown in figure 1 (c-f);
\item single top + photon production (pp $\rightarrow$ t$\gamma$), representative Feynman diagrams are shown in figure 1 (g-h);
\item top decay to up-type quark and photon in t$\bar{\text{t}}$ events (pp $\rightarrow$ t$\bar{\text{t}}$ $\rightarrow$ t$\gamma$q), the representative Feynman diagram is shown in figure 1 (i).
\end{itemize}

In the presence of tq$\gamma$ FCNC couplings, the direct single top quark production is possible in pp collisions when the photon is emitted from one of the protons. In the final state of this process, no additional particle is produced in association with the top quark. This feature can be used to differentiate this process from the SM single top quark production process where an additional jet is present in the final state. 
The direct single top quark production via tqg FCNC couplings (see figure 1(a)) is probed experimentally by the ATLAS collaboration at $\sqrt{s}=8$ TeV \cite{Aad:2015gea}.
The sensitivity of the direct single top quark production for constraining the tq$\gamma$ couplings will be discussed   in detail in  latter sections.  
\\
\\
The pp $\rightarrow$ t+jet process can occur via two ways:  the photon is in the initial state (see figure 1 (c,d)) and a photon is in intermediate state (see figure 1 (e,f)). Due to the contribution of the various particles with different PDF and the presence of the QED and QCD coupling constants, all diagrams play a role in the total cross section. 
The final state of the pp $\rightarrow$ t+jet process is almost similar to the one from SM single top quark production.
The t+jet final state is probed by the CMS collaboration at $\sqrt{s}=8$ TeV to search for tqg FCNC signatures \cite{CMS:2014ffa}. This process also has not been used to search for tq$\gamma$ FCNC interactions experimentally.
\\
\\
In addition to the photon induced signatures, the anomalous tq$\gamma$ FCNC couplings leads to the anomalous single top quark production in association with a photon and to decays of top quark to an up-type quark and a photon.
These two processes, pp $\rightarrow$ t$\gamma$ and pp $\rightarrow$ t$\bar{\text{t}}$ $\rightarrow$ t$\gamma$q, 
have little SM background compared to pp $\rightarrow$ t and pp $\rightarrow$ t+jet processes because of the photon in the final state. The pp $\rightarrow$ t$\gamma$ process is employed by the CMS collaboration and the most stringent limit  on the anomalous tq$\gamma$ couplings is set \cite{Khachatryan:2015att}.

\subsection{Numerical results}
\label{numeric}
The cross sections of direct single top quark production via tq$\gamma$ FCNC interactions for elastic and inelastic photon emission as a function of the anomalous couplings $\kappa_{\text{u}\gamma}$ and $\kappa_{\text{c}\gamma}$ for $\sqrt{s}=$ 8 and 13 TeV  are summarized in table \ref{tCS}. The contribution of the tu$\gamma$ anomalous coupling to the total cross section is  larger than the one of the tc$\gamma$ coupling since the u quark PDF has the dominant distribution (for $x$ values above 0.1).  The inelastic cross section is approximately three times larger than the elastic one for a given anomalous coupling. However, the contribution of the elastic interaction to the total cross section is not negligible. 

\begin{table}[ht]
\centering
\begin{tabular}{llll}
\hline
Photon emission  & PDF set    & $\sqrt{s}$ & Cross section (pb)     \\
\hline
\hline
Elastic  & EPA    & 8 TeV  & 225 $\kappa_{\text{u}\gamma}^2$ + 68  $\kappa_{\text{c}\gamma}^2$    \\
Elastic  & EPA    & 13 TeV  & 366 $\kappa_{\text{u}\gamma}^2$ + 140  $\kappa_{\text{c}\gamma}^2$     \\
\hline
Inelastic  & MRST\_QED    & 8 TeV  &  829 $\kappa_{\text{u}\gamma}^2$ +  267 $\kappa_{\text{c}\gamma}^2$     \\
Inelastic  & CTEQ\_QED    & 8 TeV  &  505 $\kappa_{\text{u}\gamma}^2$ +   133 $\kappa_{\text{c}\gamma}^2$     \\
Inelastic  & NNPDF\_QED    & 8 TeV  & 687 $\kappa_{\text{u}\gamma}^2$ + 294  $\kappa_{\text{c}\gamma}^2$     \\
\hline
Inelastic  & MRST\_QED  & 13 TeV  &  1392 $\kappa_{\text{u}\gamma}^2$ +  546  $\kappa_{\text{c}\gamma}^2$     \\
Inelastic  & CTEQ\_QED   & 13 TeV  & 905 $\kappa_{\text{u}\gamma}^2$ +  295 $\kappa_{\text{c}\gamma}^2$     \\
Inelastic  & NNPDF\_QED    & 13 TeV  &  1082 $\kappa_{\text{u}\gamma}^2$ + 546  $\kappa_{\text{c}\gamma}^2$     \\
\hline
\end{tabular}
\caption{The total cross section of pp $\rightarrow$ t for both elastic and inelastic photon emission as a function of the tq$\gamma$ FCNC couplings. The cross sections are calculated at the center of mass energy of 8 and  13 TeV, and for various PDF sets.}
\label{tCS}
\end{table}

The cross section of the pp $\rightarrow$ t+jet process is calculated for inelastic photon emission with the requirement  on the transverse momentum of the jet ($p_T^{\text{jet}}>10$ GeV) for  8 and 13 TeV. Results are presented in table \ref{tjetCS} (first row). The pp $\rightarrow$ t+jet process contains one extra QCD or QED vertex compared to the pp $\rightarrow$ t process which leads to a reduction in the cross section value. In table \ref{tjetCS} (second row) the total cross section of pp $\rightarrow$ t$\gamma$ as a function of the anomalous tq$\gamma$ FCNC couplings at 8 and 13 TeV are shown. The cross sections are calculated with the requirement $p_T^{\gamma}>10$ GeV.
  

\begin{table}[ht]
\centering
\begin{tabular}{lll}
\hline
Process    & $\sqrt{s}$ & Cross section (pb)     \\
\hline
\hline
\multirow{2}{*}{pp $\rightarrow$ t+jet}     & 8 TeV  & 519 $\kappa_{\text{u}\gamma}^2$ +  227 $\kappa_{\text{c}\gamma}^2$     \\
    & 13 TeV  &  998 $\kappa_{\text{u}\gamma}^2$ + 501  $\kappa_{\text{c}\gamma}^2$     \\
\hline
\multirow{2}{*}{pp $\rightarrow$ t$\gamma$}    & 8 TeV  & 96 $\kappa_{\text{u}\gamma}^2$ +  11 $\kappa_{\text{c}\gamma}^2$     \\
    & 13 TeV  &  235 $\kappa_{\text{u}\gamma}^2$ + 36  $\kappa_{\text{c}\gamma}^2$     \\
\hline
\end{tabular}
\caption{The total cross section of pp $\rightarrow$ t+jet and pp $\rightarrow$ t$\gamma$ as a function of the tq$\gamma$ FCNC couplings. The cross sections are calculated at center of mass energies of 8 and  13 TeV requiring $p_T^{\text{jet}}>10$ GeV and $p_T^{\gamma}>10$ GeV.}
\label{tjetCS}
\end{table}  

\begin{figure}[t]
\centering 
\includegraphics[width=.48\textwidth]{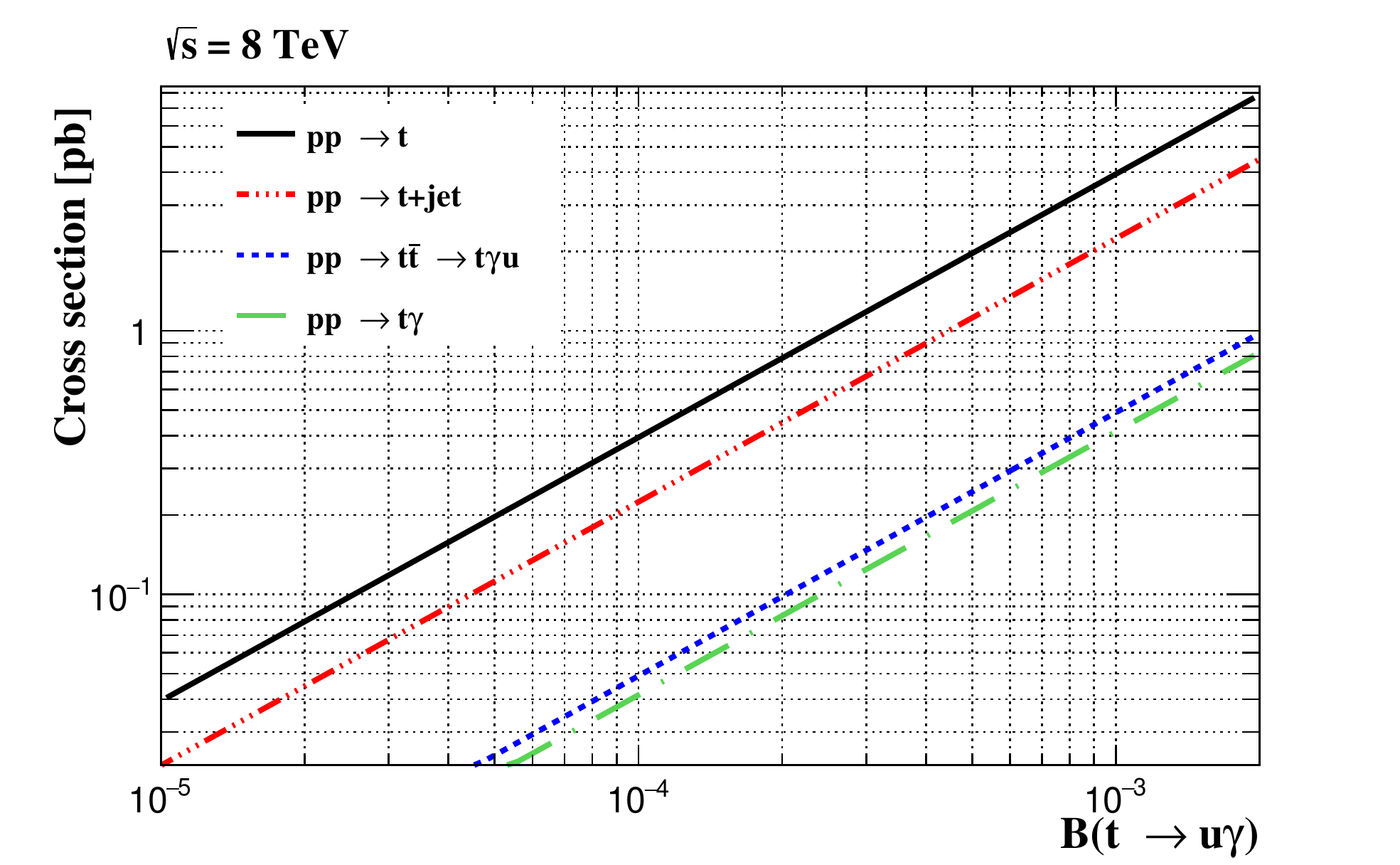}
\includegraphics[width=.48\textwidth]{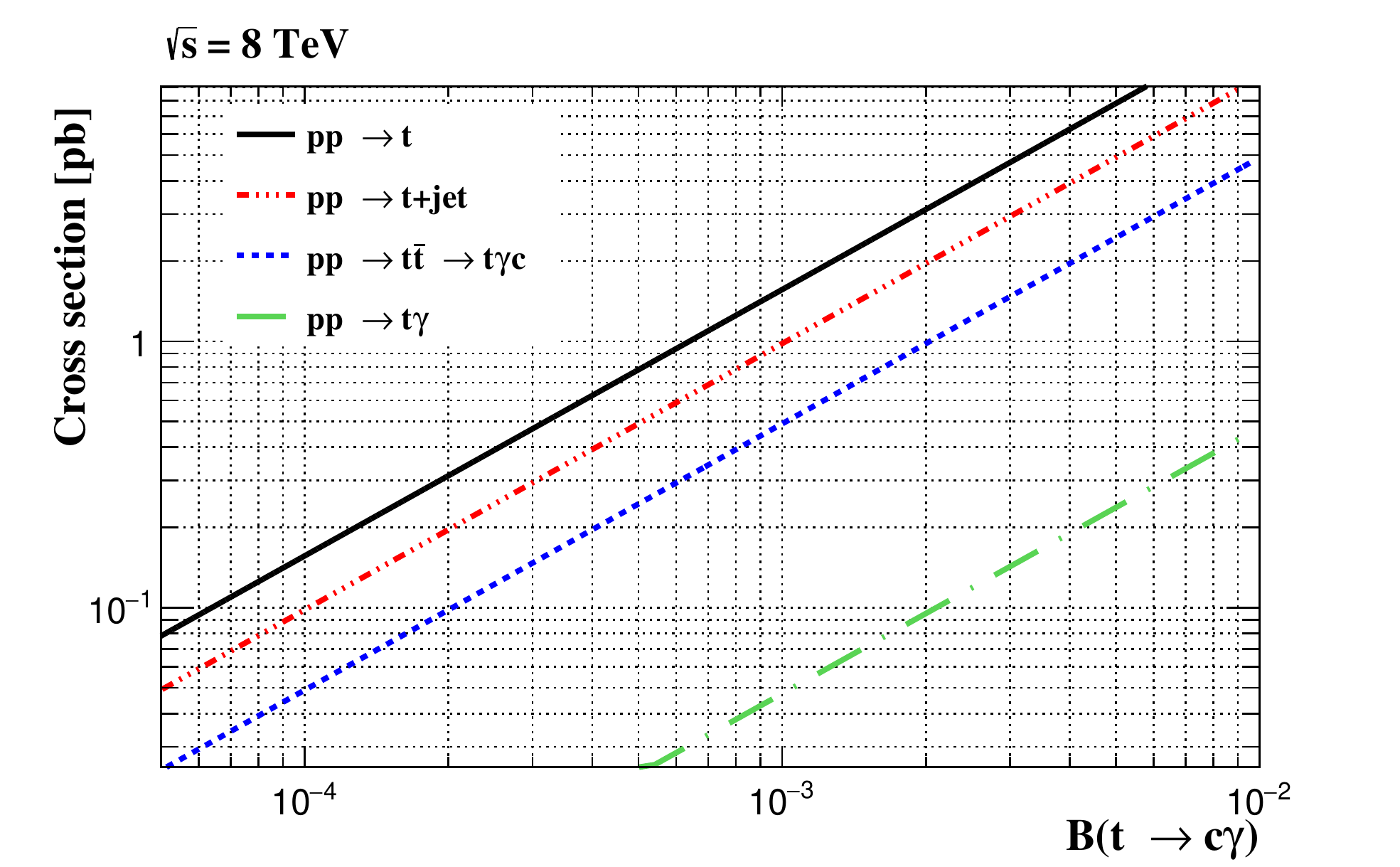}
\includegraphics[width=.48\textwidth]{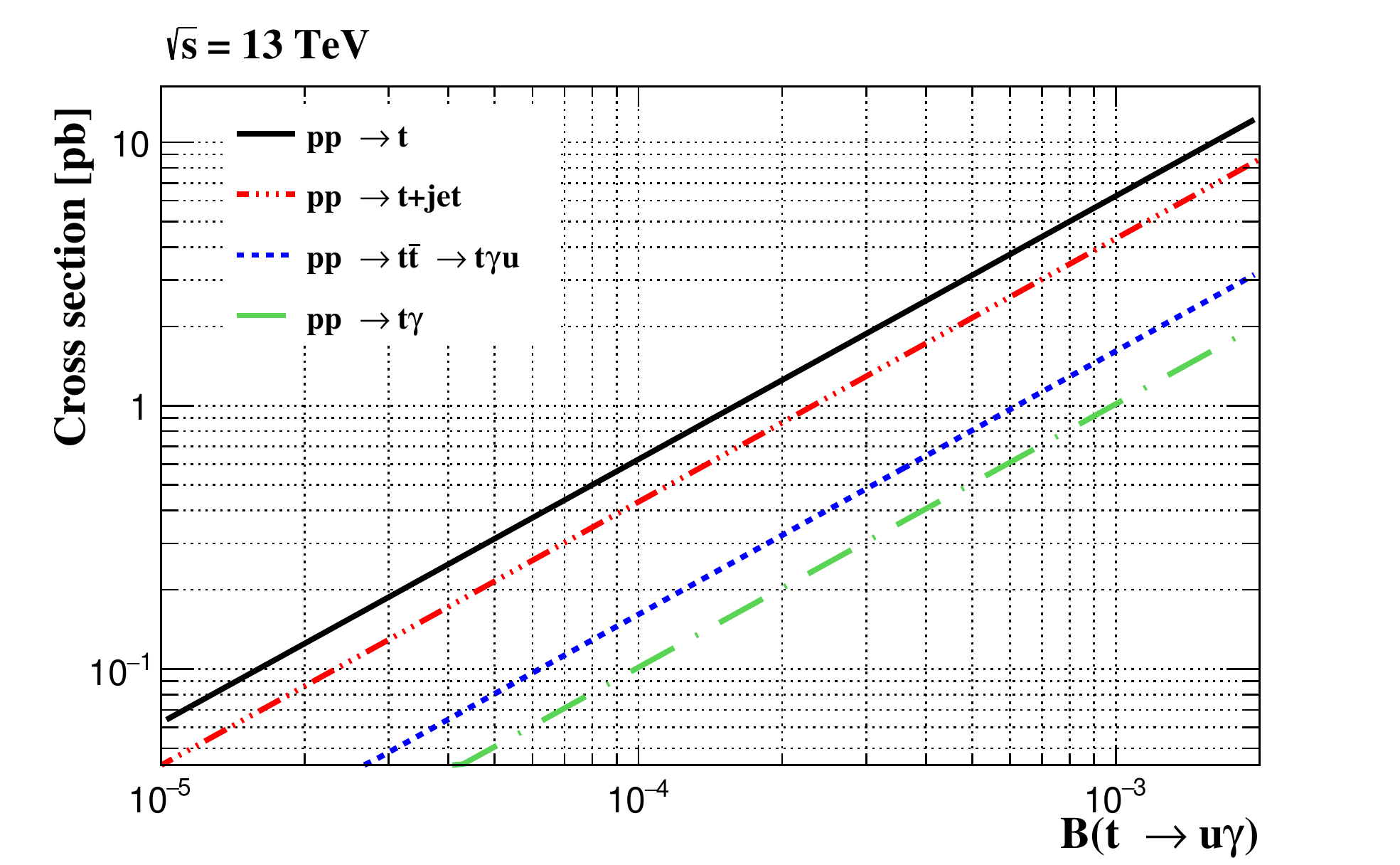}
\includegraphics[width=.48\textwidth]{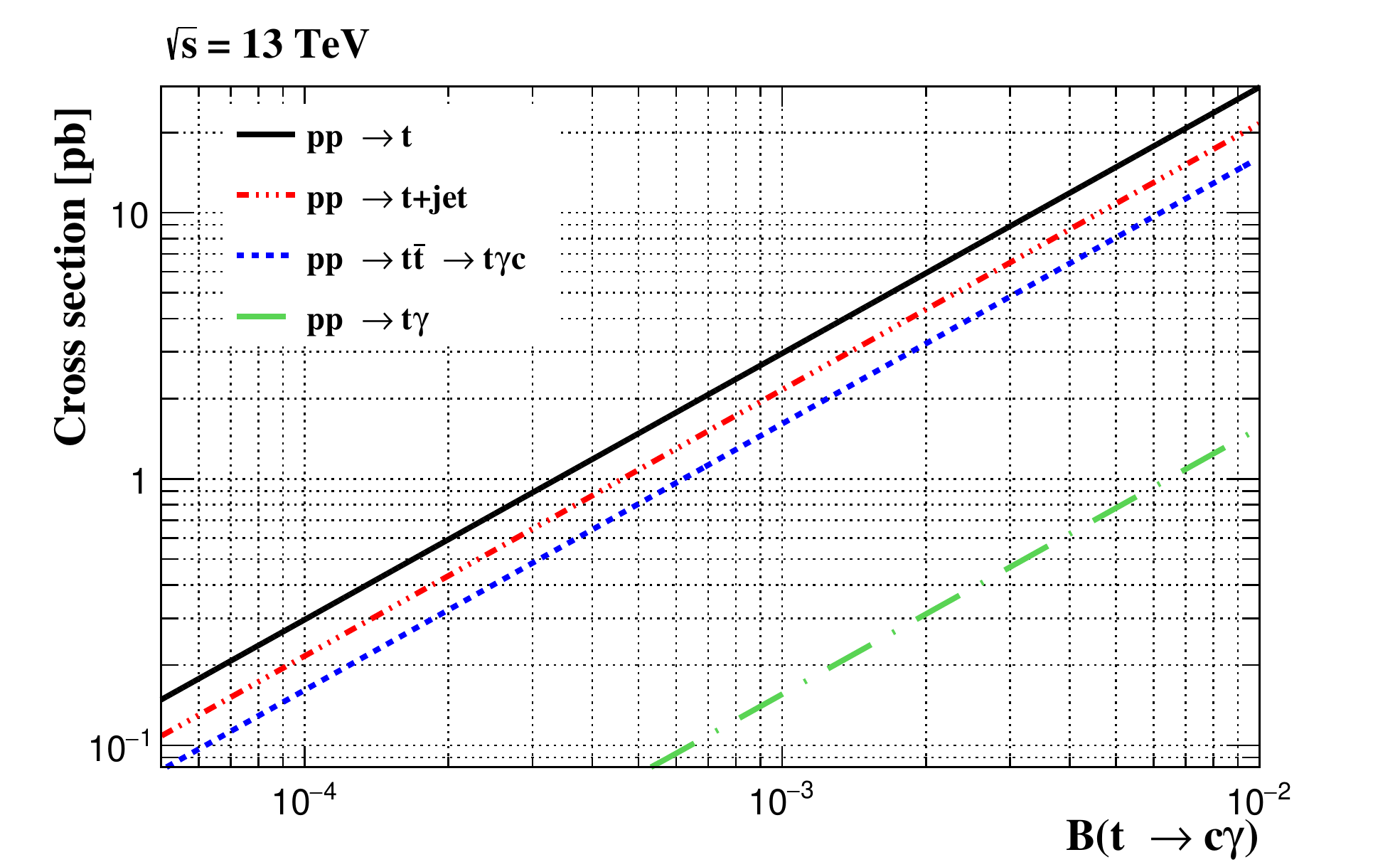}
\hfill
\caption{The total signal cross section of pp $\rightarrow$ t (elastic (EPA) + inelastic (NNPDF\_QED)), pp $\rightarrow$ t$\gamma$, pp $\rightarrow$ t+jet and pp $\rightarrow$ t$\bar{\text{t}}$ $\rightarrow$  t$\gamma$q via tu$\gamma$ (left) and t$\gamma$c (right) FCNC interactions  as a function of ${\cal B}$(t $\rightarrow$ q$\gamma)$ at 8 TeV (top) and 13 TeV (bottom).}
\label{crossSEC}
\end{figure}

Finally the cross section of the anomalous top quark decays in t$\bar{\text{t}}$ events can be written as:  
\begin{eqnarray}
\label{equBR}
&&\sigma^{\text{tq}\gamma} (\text{pp} \rightarrow \text{tq}\gamma) = \sigma^{\text{SM}} (\text{pp} \rightarrow \text{t}\bar{\text{t}}) \times  {\cal B}(\text{t} \rightarrow \text{q}\gamma) \times 2, \\
 &&\sigma^{\text{tq}\gamma}_{8 \text{TeV}} (\text{pp} \rightarrow \text{t}\gamma \text{q}) = 112.7 |\kappa_{\text{q}\gamma}|^2  , \ \  \sigma^{\text{tq}\gamma}_{13 \text{TeV}} (\text{pp} \rightarrow \text{t}\gamma \text{q}) = 370.76 |\kappa_{\text{q}\gamma}|^2,   
\end{eqnarray}
where
\begin{eqnarray}
{\cal B}(\text{t}\rightarrow \text{q}\gamma) &=& \frac{\Gamma (\text{t} \rightarrow \text{q}\gamma)}{\Gamma (\text{t} \rightarrow \text{Wb})} = 0.230 |\kappa_{\text{q}\gamma}|^2,  \\
\Gamma (\text{t} \rightarrow \text{q}\gamma) &=& \frac{\alpha}{4} m_{\text{t}} |\kappa_{\text{q}\gamma}|^2, \\
\Gamma (\text{t} \rightarrow \text{Wb}) &=& \frac{\alpha}{16 s_w^2} |V_{\text{tb}}|^2 \frac{m_{\text{t}}^3}{m_{\text{W}}^2} \left[1-3\frac{m_{\text{W}}^4}{m_{\text{t}}^4} +2\frac{m_{\text{W}}^6}{m_{\text{t}}^6}\right],  
\end{eqnarray}
in which $m_{\text{t}}=173$ GeV, $s_W^2=0.234$ and $\alpha= 1/128.92$. In equation 5, we have used the NNLO QCD values of 245 pb and 806 pb for t$\bar{\text{t}}$ cross section at $\sqrt{s}=$ 8 and 13 TeV, respectively \cite{Czakon:2013goa}.      
\\
\\
In figure \ref{crossSEC}, the cross sections of all four mentioned channels for probing tq$\gamma$ FCNC interactions are shown as a function of the branching fraction of the t $\rightarrow$ q$\gamma$ anomalous decays at 8 and 13 TeV.  One can see that the photon induced direct single top quark production has the highest cross section for a given value for the branching fraction, followed by the pp $\rightarrow$ t+jet channel. Therefore, considering the direct single top quark production can lead to  a significant improvement in the sensitivity of the searches for the $tq\gamma$ FCNC at the LHC. 
\\
\\
In the presence of the tu$\gamma$ FCNC coupling ($\kappa_{c\gamma}=0$), the pp $\rightarrow$ t$\gamma$ channel and the pp $\rightarrow$ t$\bar{\text{t}}$ $\rightarrow$ t$\gamma$u channel have almost the same cross section at 8 TeV. However, the  pp $\rightarrow$ t$\bar{\text{t}}$ $\rightarrow$ t$\gamma$u cross section benefits more from the increase of the center of mass energy from 8 to 13 TeV (see figure \ref{crossSEC} (left)).
In order to exclude ${\cal B}$(t $\rightarrow$ u$\gamma)$ for values above $10^{-4}$, we need to reach the values of 40 fb, 50 fb, 200 fb and 400 fb for the upper bounds on the cross sections in pp $\rightarrow$ t$\gamma$, pp $\rightarrow$ t$\bar{\text{t}}$ $\rightarrow$ t$\gamma$u, pp $\rightarrow$ t+jet and direct single top quark channels respectively for $\sqrt{s}=8$ TeV. For $\sqrt{s}=13$ TeV, these values are 100 fb, 170 fb, 400 fb and 600 fb in pp $\rightarrow$ t$\gamma$, pp $\rightarrow$ t$\bar{\text{t}}$ $\rightarrow$ t$\gamma$u, pp $\rightarrow$ t+jet and direct single top quark channels.
\\
\\
The pp $\rightarrow$ t$\gamma$ channel has the lowest cross section among these four channels for a given value of the tc$\gamma$ FCNC coupling ($\kappa_{u\gamma}=0$). For instance, the cross section of pp $\rightarrow$ t$\gamma$ process is around 30 (20) times smaller than the cross section of direct single top quark production   at $\sqrt{s}=8$ TeV (13 TeV).
The cross section of the pp $\rightarrow$ t$\bar{\text{t}}$ $\rightarrow$ t$\gamma$c channel is the same  as pp $\rightarrow$ t$\bar{\text{t}}$ $\rightarrow$ t$\gamma$u channel and increases by a factor of $\sim$3 from 8 TeV to 13 TeV.

\section{Analysis setup}
Four signal channels, for probing the tq$\gamma$ FCNC interactions in pp collisions, were discussed in section 2.
All signal channels have single top quark in the final state. 
\\
\\
The SM backgrounds for each signal channel are different due to the presence or not of extra objects in the final state. Therefore, different event selection should be used to reach the highest sensitivity for each channel. The direct single top quark production (pp $\rightarrow$ t) and single top quark production in association with a jet (pp $\rightarrow$ t+jet) final states have been studied to search for tqg FCNC interactions by the ATLAS and CMS collaboration, respectively \cite{Aad:2015gea,CMS:2014ffa}. No excess over the predicted SM background has been observed  and  the FCNC parameters are constrained consequently.
The results of the pp $\rightarrow$ t and pp $\rightarrow$ t+jet can be reinterpreted to constrain the tq$\gamma$ FCNC interactions.
The main purpose of this paper is   to study the sensitivity of the direct single top quark production channel for probing the FCNC tq$\gamma$ interactions. Therefore, we will focus on the direct single top quark production and add the pp$\rightarrow$ t+jet, pp $\rightarrow$ t$\gamma$ and pp $\rightarrow$ t$\bar{\text{t}}$ $\rightarrow$  t$\gamma$q contribution to the signal region to increase the search sensitivity. 

\subsection{Experimental input}
\label{EI}
Since the hadronic decay modes of the top quark suffer from large background from QCD processes, we focus on its leptonic (electron or muon) decay channels.  Events from q$\gamma$ $\rightarrow$ t $\rightarrow$ Wb $\rightarrow$ $\Plepton \nu$b are characterized by an isolated electron or muon, exactly one jet from the b-quark hadronization and missing transverse energy ($E_T^{miss}$) due to the presence of an undetected neutrino in the final state. 
In order to use the experimental published results to evaluate the sensitivity of direct single top quark production channel for probing the tq$\gamma$ interactions we follow closely the analysis strategy presented in \cite{Aad:2015gea} by the ATLAS collaboration.  In the ATLAS analysis, a search is performed for direct single top quark production via tqg FCNC interactions using 8 TeV pp collision data, corresponding to an integrated luminosity of 20.3 fb$^{-1}$.   
\\
\\
The events in data and Monte-Carlo simulations are selected as follows:
\begin{itemize}
\item exactly one electron candidate with $p_T$ greater than 25 GeV  and $|\eta|<2.47$ (the calorimeter barrel-endcap transition region of the ATLAS detector, $1.37<|\eta|<1.52$, is excluded), or exactly one muon with $p_T>25$ GeV and $|\eta|<2.5$. In addition, identification and isolation criteria are applied on the electron and muon candidates (see details in \cite{Aad:2015gea});
\item jets are reconstructed using the anti-$k_t$ algorithm \cite{Cacciari:2008gp} with a radius parameter of 0.4 and are required to have $p_T>25$ GeV and $|\eta|<2.5$. Among the selected jet, exactly one jet, with  $p_T>30$ GeV and $|\eta|<2.5$ originating from b-quark (b-tagged jet) is required;
\item $E_T^{miss}>30$ GeV and $m_T(\text{W})>50$ GeV where  $m_T(\text{W})$ is the transverse mass of the W boson calculated as $\sqrt{2\left( p_T (\Plepton)E_T^{miss} - \vec{p}_T (\Plepton) .  \vec{E}_T^{miss}\right)}$;
\item $p_T^{\Plepton} > 90 \left(1- \frac{\pi - |\Delta \phi (\Plepton, \text{b-jet})}{\pi -2} \right)$ GeV. 
\end{itemize}

Various SM processes with a similar signature can mimic the new physics signal events. The most important background sources for this analysis are from SM top-quark  (single top and t$\bar{\text{t}}$), W+jets, Z+jets and multi-jet production. 
The contribution of the SM backgrounds is estimated with MC simulated samples normalized to the total integrated luminosity except for multi-jet background. Data driven methods are used to estimate the normalization of the multi-jet background while the shape is established from simulated samples (more details can be found in \cite{Aad:2015gea}). After event selection and background estimation procedure, ATLAS data are well described by  the SM prediction and no sign of new physics signal was observed.

\subsection{Signal simulation}
Signal events are generated using the {\sc MadGraph\_}a{\sc mc@NLO} event generator at 8 TeV as described in section 2. Due to the fact that the kinematic distributions of the final state particles are independent of the strength of the FCNC couplings, signal samples are produced with an arbitrary assumption for $\kappa_{\text{u}\gamma}$ and $\kappa_{\text{c}\gamma}$. 
For each signal process, three samples are generated: [$\kappa_{\text{u}\gamma}$ = 0.1, $\kappa_{\text{c}\gamma}$ = 0], [$\kappa_{\text{u}\gamma}$ = 0, $\kappa_{\text{c}\gamma}$ = 0.1] and [$\kappa_{\text{u}\gamma}$ = 0.1, $\kappa_{\text{c}\gamma}$ = 0.1]. For the pp $\rightarrow$ t$\gamma$ and pp $\rightarrow$ t+jet processes, thresholds are set on the photon and jet transverse momenta as discussed in section \ref{numeric}. 
The branching fraction of the top quark to W boson and b quark is assumed to be 100\% at the generator level. Signal events from tau decays to electron or muon are also included in signal samples.
The NNPDF\_QED  is used as the nominal PDF \cite{Ball:2013hta}.
\\
\\
Generated events are passed through {\sc Pythia} 8 \cite{Sjostrand:2014zea} for the parton showering and hadronisation.
In order to simulate the detector effects, {\sc Delphes} \cite{delphes} simulator is employed. The official ATLAS detector card in {\sc Delphes} is used, except for the b-tag efficiency (57\%) and miss-tag rate (0.2\%), as reported in ref. \cite{Aad:2015gea}. 
\\
\\
The signal selection efficiency for direct top quark production via tqg FCNC interaction is estimated to be 3.1\% by the ATLAS collaboration \cite{Aad:2015gea}. In table \ref{seff},  results of our simulation for signal selection efficiencies are presented.  In the first row, the signal selection efficiency for the direct top quark production via tqg FCNC interactions is shown. Our calculated efficiency is close to the ATLAS one although slightly lower. This difference can be due to some phenomena that we have not considered in our simulation like pileup and trigger effects. This can also come from  the energy scale, energy resolution, identification and isolation differences for electron, muon and jets between full simulated ATLAS samples and our fast simulated samples.
In order to account for the difference between the {\sc Delphes} simulated samples and the ATLAS samples we normalize the signal selection efficiencies to the values extracted from the tqg FCNC samples.
\begin{equation}
\epsilon_{\text{tq}\gamma} = \epsilon^{\text{DELPHES}}_{\text{tq}\gamma} \times \frac{\epsilon^{\text{ATLAS}}_{\text{tqg}}}{\epsilon^{\text{DELPHES}}_{\text{tqg}}}
\end{equation}
\begin{table}[h]
\centering
\begin{tabular}{llcccc}
\hline
Process           & FCNC vertex                                                       & $\epsilon (\kappa_{\text{u}\gamma}$) & $\epsilon (\kappa_{\text{c}\gamma})$ & $\epsilon (\kappa_{\text{u}\gamma} = \kappa_{\text{c}\gamma})$  & Normalized $\epsilon$ \\
\hline
\hline
pp $\rightarrow$ t & tqg                               & 1.86\%       & 2.59\%       & 2.00\%                       & 3.1\% \\
pp $\rightarrow$ t  & tq$\gamma$                              & 2.50\%        & 2.30\%        & 2.40\%                     &3.72\% \\
pp $\rightarrow$ t+jet & tq$\gamma$                           & 1.16\%       & 0.95\%       & 1.11\%                    & 1.79\%\\
pp $\rightarrow$ t$\gamma$  & tq$\gamma$                         & 1.88\%       & 2.01\%       & 1.86\%                    & 2.88\%\\
pp $\rightarrow$ t$\bar{\text{t}}$ $\rightarrow$  t$\gamma$q & tq$\gamma$  & 0.27\%       & 0.31\%       & 0.31\%                     &0.48\%\\
\hline
\end{tabular}
\caption{Signal selection efficiency for various signal sample with tq$\gamma$ and tq$g$ FCNC couplings. The last column gives the signal selection efficiency normalized to the $\frac{\epsilon^{\text{ATLAS}}_{\text{tqg}}}{\epsilon^{\text{DELPHES}}_{\text{tqg}}}$. }
\label{seff}
\end{table}
\\
\\
It can be seen in table \ref{seff} that the selection is slightly more efficient for the photon initiated production compared to gluon initiated one. The presence of an extra jet in pp $\rightarrow$ t+jet and pp $\rightarrow$ t$\bar{\text{t}}$ $\rightarrow$  t$\gamma$u  process decreases the signal selection efficiency for these processes compared to direct single top quark production.   
The signal selection efficiency for the pp $\rightarrow$ t$\gamma$ process is comparable with pp $\rightarrow$ t process since no selection is required on the photon in final state.
The signal selection efficiencies are reported for different coupling assumptions.
Various combinations of $\kappa_{\text{u}\gamma}$ and $\kappa_{\text{c}\gamma}$ lead to small changes in signal selection efficiencies due to the different PDF of the up quark and charm quark. However, it is observed that the kinematic distributions of the final state particles have almost the same behavior  for various combination of the $\kappa_{\text{u}\gamma}$ and $\kappa_{\text{c}\gamma}$ anomalous couplings.

\section{Limits}

\begin{figure}[th]
\centering 
\subfloat[][]{
\includegraphics[width=.48\textwidth]{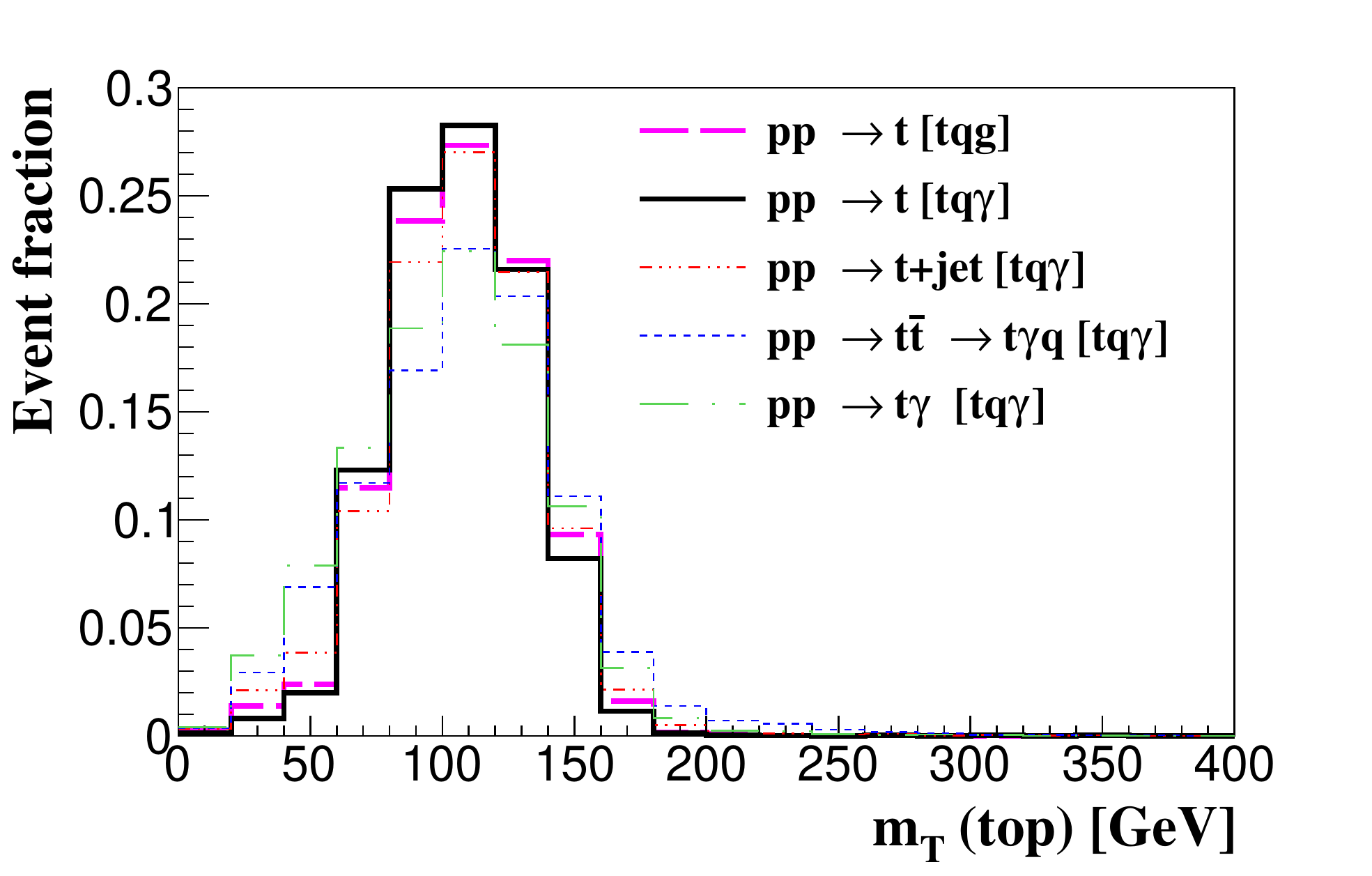}{}}
\subfloat[][]{
\includegraphics[width=.48\textwidth]{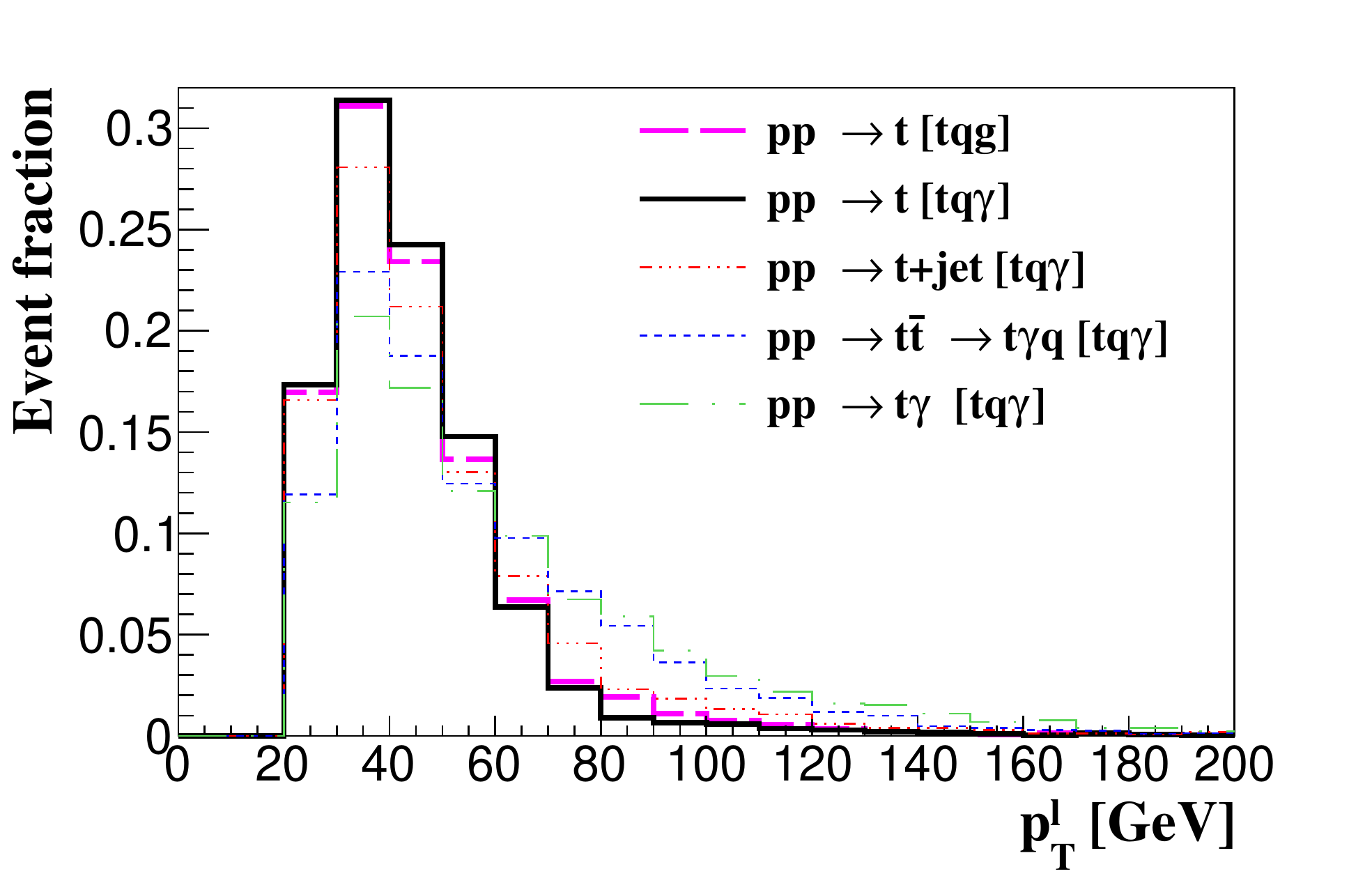}{}} \\
\subfloat[][]{
\includegraphics[width=.48\textwidth]{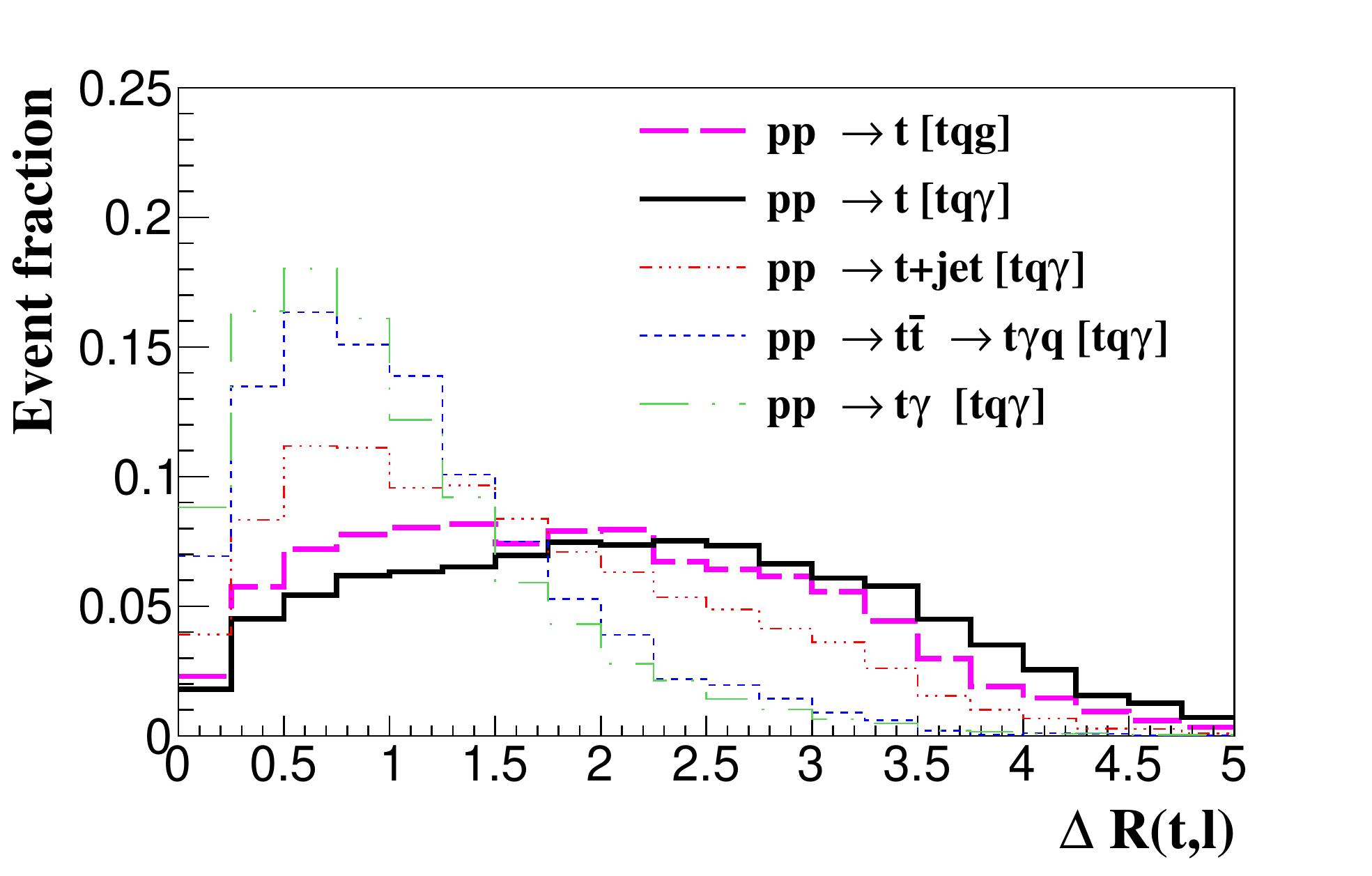}{}}
\subfloat[][]{
\includegraphics[width=.48\textwidth]{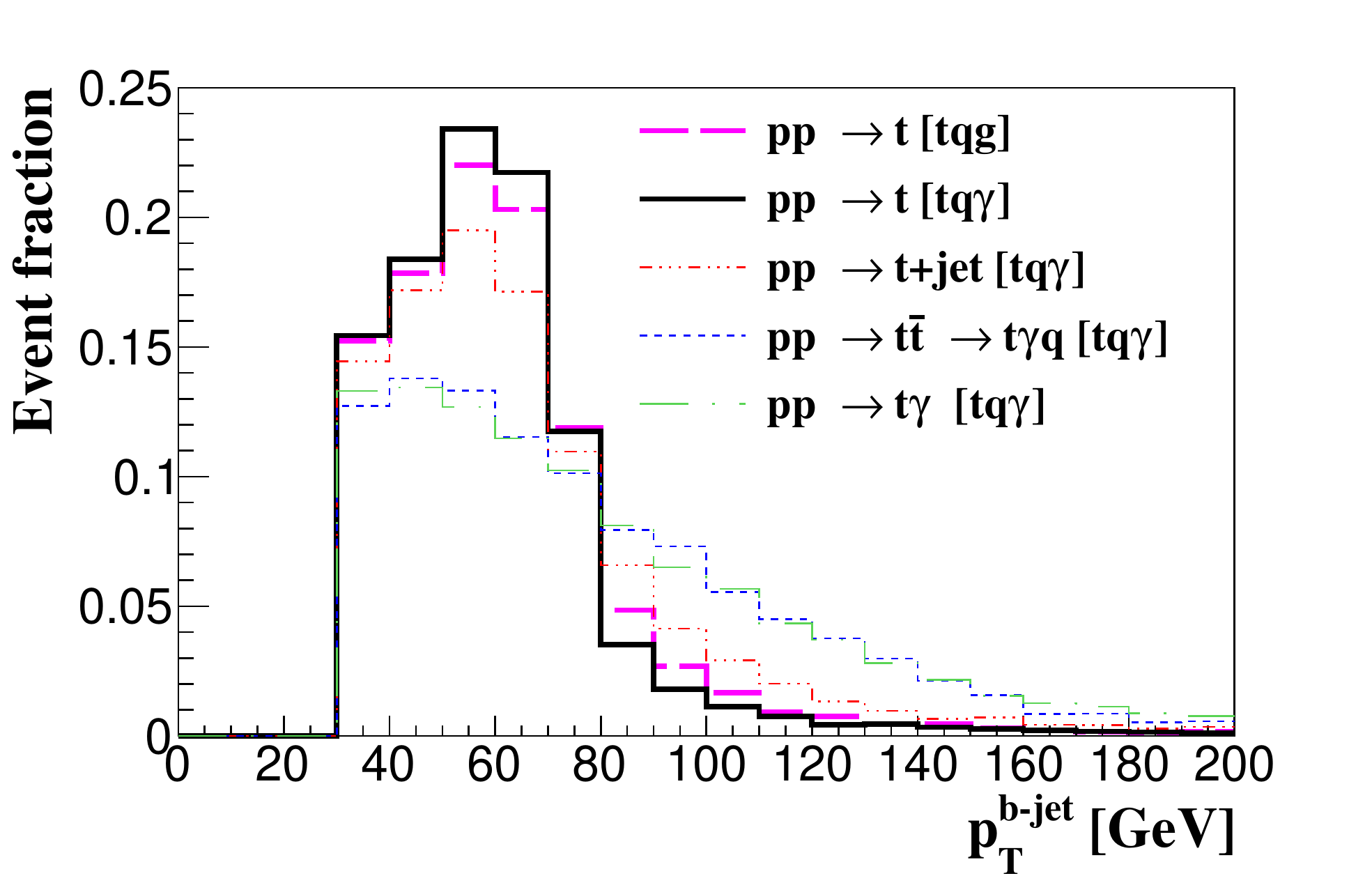}{}}
\hfill
\caption{Distributions of four important discriminating variables: (a) the top quark transverse mass (b) the lepton transverse momentum (c) the $\Delta R$ between the lepton and the reconstructed top quark (d) the b-jet transverse momentum for the  pp $\rightarrow$ t, pp $\rightarrow$ t+jet, pp $\rightarrow$ t$\gamma$ and pp $\rightarrow$ t$\bar{\text{t}}$ $\rightarrow$  t$\gamma$q. }
\label{NNvar}
\end{figure}

\begin{figure}[t]
\centering 
\includegraphics[width=.48\textwidth]{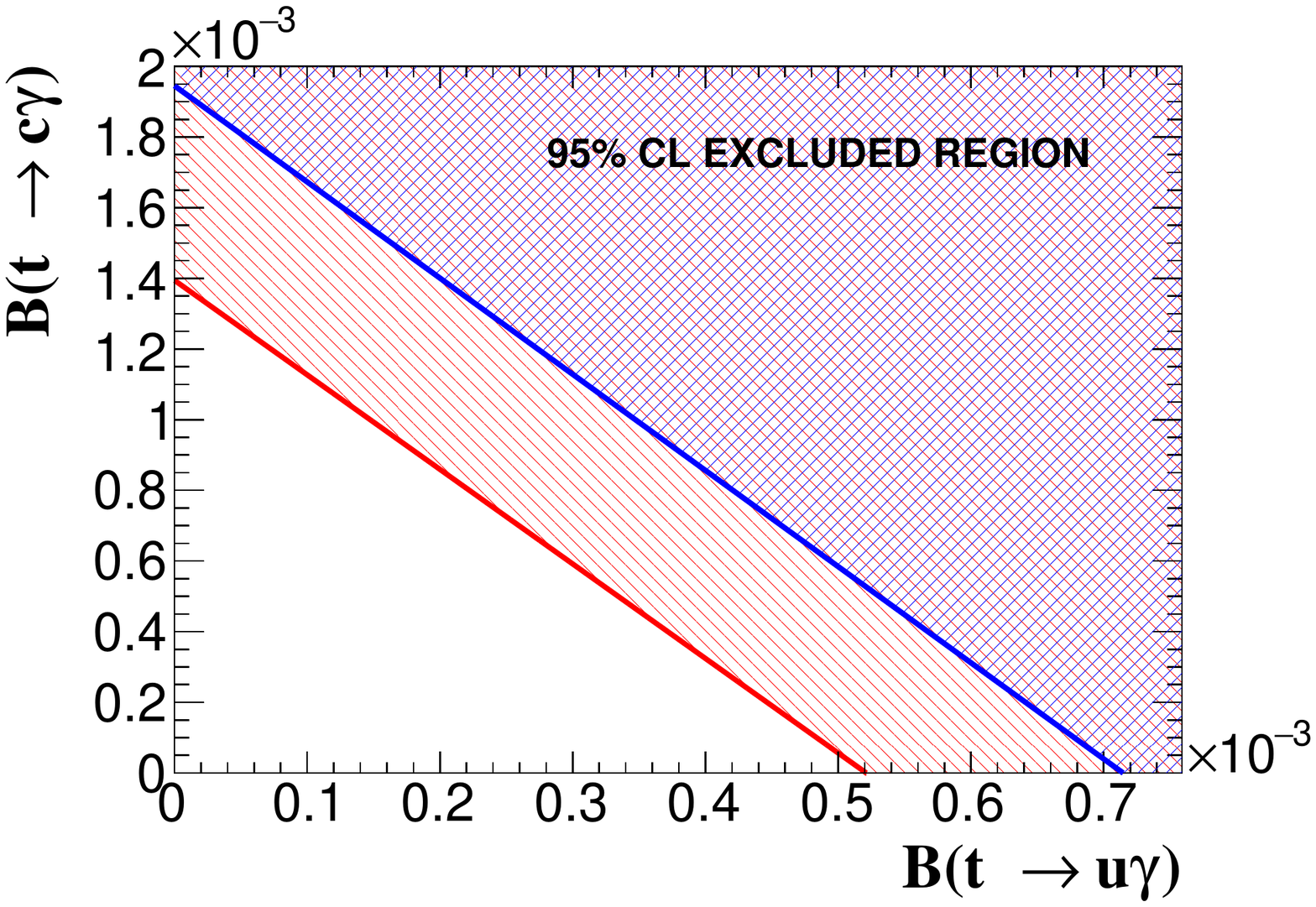}
\includegraphics[width=.48\textwidth]{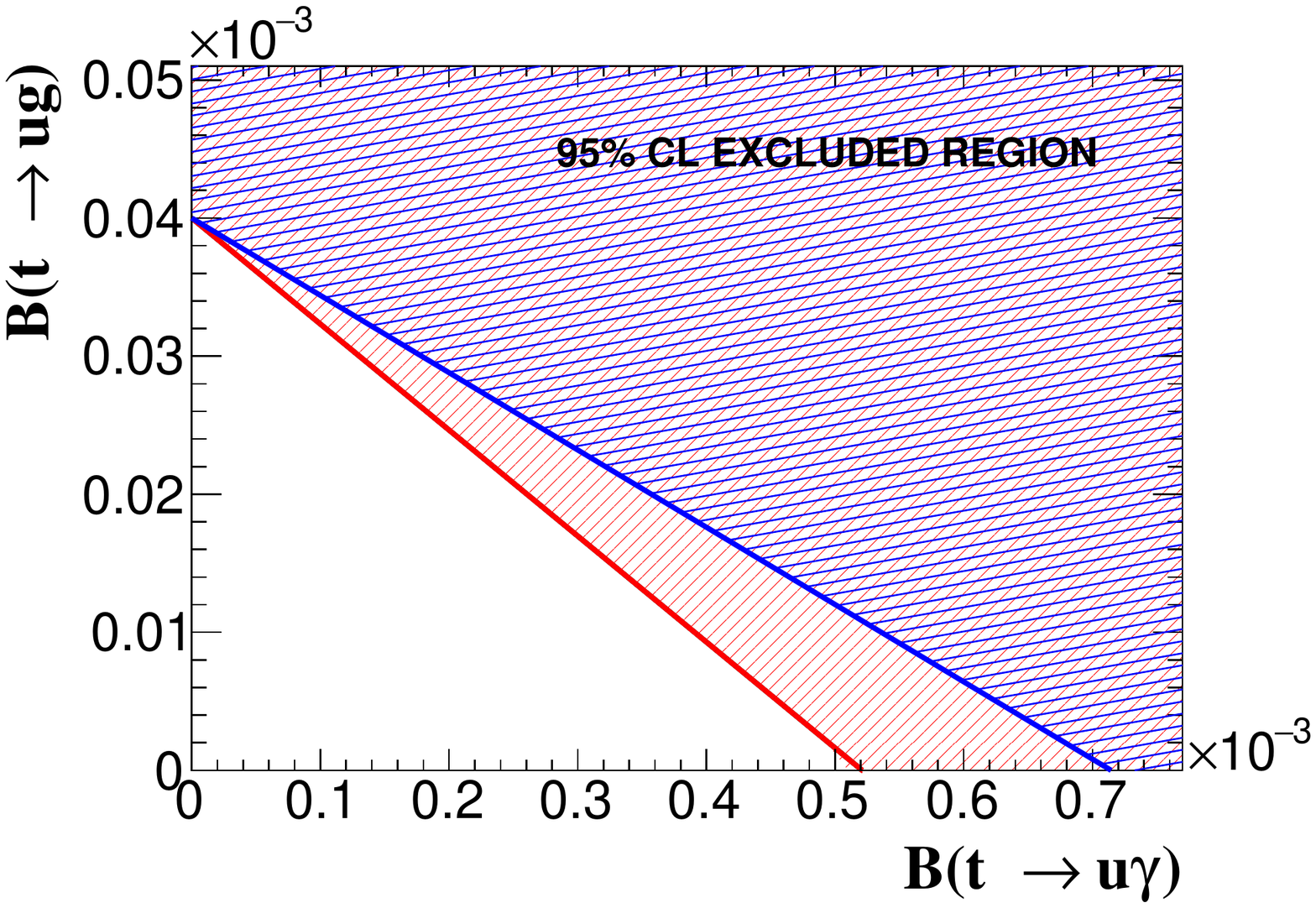}
\hfill
\caption{Excluded region at 95\% CL on the branching fraction plane of (${\cal B}$(t $\rightarrow$ c$\gamma)$, ${\cal B}$(t $\rightarrow$ u$\gamma)$) (left) and (${\cal B}$(t $\rightarrow$ ug), ${\cal B}$(t $\rightarrow$ u$\gamma)$) (right). The blue hatched regions show the branching fraction values excluded when considering the pp $\rightarrow$ t process and the red hatched regions show the branching fraction values excluded when considering all signal channels. }
\label{limitplots}
\end{figure}

In the search for direct single top quark production performed by the ATLAS collaboration, multivariate  analysis  techniques  have  been  used  to separate SM background events from signal candidates. To find a single powerful discriminator, the following variables are combined in a neural network classifier \cite{Feindt:2006pm}: transverse masses of the reconstructed top quark and W boson, transverse momenta of the lepton, the b-jet and the W boson, charge of the lepton, pseudorapidities of the lepton and the reconstructed top quark, $\Delta R$(t,$\Plepton$),  $\Delta R$(t,b-jet), $\Delta \phi$(t,W) and cos$\theta$($\Plepton$, b-jet). A binned maximum likelihood fit is performed to the neural network output distribution in the signal region. As no clear evidence of signal is observed, 95\% confidence level upper limit  on the anomalous FCNC single top quark production times the t $\rightarrow$ Wb branching fraction is set to 3.4 pb, while $2.9^{+1.9}_{-1.2}$ pb is expected \cite{Aad:2015gea}.
\\
\\
Although using the multivariate analysis technique provides better sensitivity for experimental searches, it leads to more complicated result interpretations for phenomenologists. The upper limit in \cite{Aad:2015gea} is obtained by analyzing the shape of the neural network output.
Therefore, one should make sure that the various distributions of understudy signals are similar to the ones of the ATLAS experimental search in order to use their experimental results.
\\
\\
Figure \ref{NNvar} presents the distributions of the four most important discriminating variables: the transverse mass of the reconstructed top quark, the transverse momentum of the lepton, $\Delta R$(t,$\Plepton$) and the transverse momentum of the b-jet. The direct single top quark production due to the anomalous FCNC tqg (signal assumed in \cite{Aad:2015gea}) and tq$\gamma$ (the photon initiated signal) are compared in figure \ref{NNvar}. These two channels have very similar distributions for the variables which are important in the neural network training. Therefore, we can use the upper bound obtained on the direct single top quark production cross section via tqg FCNC interactions to constraint the tq$\gamma$ signal strengths.
\\
\\
In addition to the main signal channel (pp $\rightarrow$ t), the distribution for other complementary signal channels are also shown in figure    \ref{NNvar}.
One can see that the pp $\rightarrow$ t+jet process has almost the same behavior as pp $\rightarrow$ t, while other channels behave differently. 
Since the important neural network variables are different for processes with an extra photon in the final state and for direct single top quark production, the observed limit estimation should be redone with the shape of all signal processes added. On the other hand, the contributions of the  pp $\rightarrow$ t, pp$\rightarrow$ t+jet, pp $\rightarrow$ t$\gamma$ and pp $\rightarrow$ t$\gamma$q to the total signal cross section are 73\% (71.5\%), 20\% (23.4\%), 1\% (3.1\%) and 6\% (2\%), respectively, for the tu$\gamma$ (tc$\gamma$) FCNC coupling. Therefore, the change in the total signal shape when adding the pp $\rightarrow$ t$\gamma$ and pp $\rightarrow$ t$\gamma$q contributions is negligible and we only consider their contributions in the total number of signal events in our limit calculation.
\\
\\
The observed upper limit reported by the ATLAS collaboration is used to constrain the anomalous top quark branching fraction to u/c quark and a photon. In table \ref{FCNCresults}, results are presented when only direct single top quark production is considered as signal and when all tq$\gamma$ signals with single top quark in the final state are added. These results are also given in figure \ref{limitplots} (left).
In the ATLAS analysis, the upper limits are set on the top quark branching fraction  FCNC decays, ${\cal B}$(t $\rightarrow$  ug)  $<4.0 \times 10^{-5} $ and ${\cal B}$(t $\rightarrow$ cg)  $<20 \times 10^{-5}$.
Due to the use of the experimental results with the same final state, the upper limits obtained in our study can be combined with the ATLAS results and exclusion in the tqg and tq$\gamma$ branching fraction plane can be obtained, as shown in figure \ref{limitplots} (right). 
\begin{table}[h]
\centering
\begin{tabular}{lcccc}
\hline
 & $\kappa_{\text{u}\gamma}$ & ${\cal B}$(t $\rightarrow$ u$\gamma)$    & $\kappa_{\text{c}\gamma}$ & ${\cal B}$(t $\rightarrow$ c$\gamma)$      \\
\hline
\hline
pp $\rightarrow$ t channel & 0.056 & 0.07\%    &  0.092 &     0.19\%   \\
All signal channels & 0.048\    &  0.05\% & 0.078  &  0.14\%  \\
\hline
\end{tabular}
\caption{The upper limit at 95\% confidence level on the tq$\gamma$ FCNC couplings and the branching fraction of top quark FCNC decays through pp $\rightarrow$ t only process and the combination of pp $\rightarrow$ t, pp $\rightarrow$ t+jet, pp $\rightarrow$ t$\gamma$ and pp $\rightarrow$ t$\bar{\text{t}}$ $\rightarrow$  t$\gamma$q processes. }
\label{FCNCresults}
\end{table}
\\
\\
One of the most important uncertainty sources for the pp $\rightarrow$ t process  is the PDF uncertainty especially originated from the photon PDF. The PDF uncertainty has been computed using {\sc SysCalc} package inside the {\sc MadGraph\_}a{\sc mc@NLO}. With the NNPDF\_QED set a variation of the cross section for the pp $\rightarrow$ t process of 12\% is obtained. Although the value of the PDF uncertainty calculated from the NNPDF\_QED set is reasonable we have observed large variations on the predicted cross sections of pp $\rightarrow$ t process due to the use of different PDFs especially between CTEQ\_QED and MRST\_QED (see table \ref{tCS}). 
This issue needs to be discussed with the related PDF collaborations and a correct uncertainty should be assigned to the signal cross section when this channel is used by experimental collaborations.
For our study, we have ignored the potential different values for the signal systematic uncertainties between the gluon initiated and photon initiated direct top production.   

\section{13 TeV projection and discussion}
\begin{figure}[t]
\centering 
\includegraphics[width=.7\textwidth]{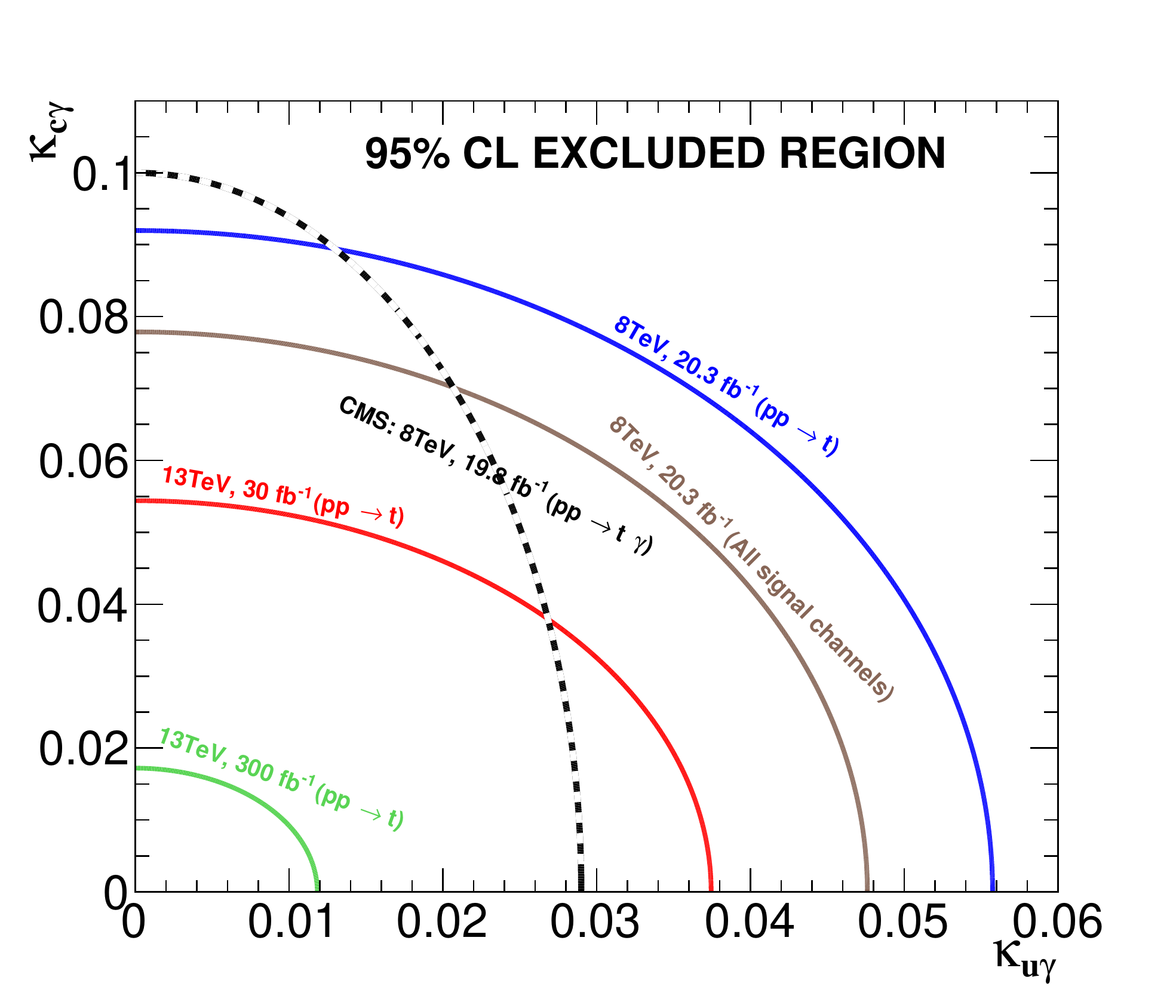}
\hfill
\caption{Excluded region at 95\% CL on the coupling constants $\kappa_{\text{u}\gamma}$  and $\kappa_{\text{c}\gamma}$ obtained via pp $\rightarrow$ t channel and all signal channels at 8 TeV (blue and brown curves). The projection results to 13 TeV are shown with red and green curves. The most stringent result obtained by the CMS collaboration via pp $\rightarrow$ t$\gamma$ at 8 TeV is shown with dashed black curve \cite{Khachatryan:2015att}.}
\label{kkplot}
\end{figure}
The ATLAS and CMS experiments have been collecting data at 13 TeV since 2015.
The LHC has reached the designed luminosity and the objective of 30 fb$^{-1}$ of data delivered to experiments for the whole of 2016 is within sight. More data will come in years 2017 and 2018. In this section we present the sensitivity of our analysis at 13 TeV and for various luminosity scenarios.
\\
\\
In order to estimate the reach of the proposed channel for probing the tq$\gamma$ FCNC coupling at higher center of mass energy we have projected the 8 TeV results obtained in this paper to 13 TeV for 30 fb$^{-1}$ and 300 fb$^{-1}$ of integrated luminosity.
In the projection procedure it is assumed that the experimental searches can reach the same upper bound as ATLAS reached at 8 TeV \cite{Aad:2015gea} on the new physics signal cross section. In figure \ref{kkplot}, the projection results are shown for 30 fb$^{-1}$ and 300 fb$^{-1}$ of integrated luminosity at 13 TeV in the $\kappa_{\text{c}\gamma}-\kappa_{\text{u}\gamma}$ plane.
The direct single top production channel can exclude the $\kappa_{\text{u}\gamma}$ and $\kappa_{\text{c}\gamma}$ couplings up to 0.037 (0.012) and 0.054 (0.017), respectively, with 30 fb$^{-1}$ (300 fb$^{-1}$) of integrated luminosity at 13 TeV.
These limits on  anomalous couplings are translated to limits on ${\cal B}$(t $\rightarrow$ u$\gamma)$ and ${\cal B}$(t $\rightarrow$ c$\gamma)$ and are summarized in table \ref{pRedresults}.
\begin{table}[h]
\centering
\begin{tabular}{lccc|c}
\hline
 & 8 TeV    & 13 TeV   & 13 TeV & 8 TeV (CMS)  \\
  & 20.3 fb$^{-1}$  & 30 fb$^{-1}$  & 300 fb$^{-1}$ & 19.8 fb$^{-1}$  \\
\hline
\hline
${\cal B}$(t $\rightarrow$ \text{u}$\gamma)$ [$\kappa_{\text{c}\gamma}=0$] & 7$\times$10$^{-4}$   &       3.2$\times$10$^{-4}$ & 3.2$\times$10$^{-5}$  &  1.3$\times$10$^{-4}$\\
${\cal B}$(t $\rightarrow$ \text{c}$\gamma)$ [$\kappa_{\text{u}\gamma}=0$] & 1.9$\times$10$^{-3}$    &       6.8$\times$10$^{-4}$ & 6.8$\times$10$^{-5}$ &  1.7$\times$10$^{-3}$ \\
\hline
\end{tabular}
\caption{The upper limit at 95\% confidence level on the branching fraction of top quark FCNC decays to u/c quark and a photon through pp $\rightarrow$ t only process at 8 and 13 TeV for various amount of data. In the last column the most stringent limits obtained by the CMS collaboration are shown \cite{Khachatryan:2015att}.}
\label{pRedresults}
\end{table}

In addition to the projection results, the limits obtained in this paper and the most stringent experimental bounds to date on the tq$\gamma$ couplings obtained in \cite{Khachatryan:2015att} by the CMS collaboration are shown in the figure.
The limits obtained by the CMS collaboration on the coupling strengths are corrected for the differences in the parameter definition in their Lagrangian.
Although the pp $\rightarrow$ t$\gamma$ channel which is used by the CMS collaboration has better sensitivity at 8 TeV for the $\kappa_{\text{u}\gamma}$ coupling (because of the clean signature and higher rejection of backgrounds) the pp $\rightarrow$ t channel allows to improve the sensitivity for the $\kappa_{\text{c}\gamma}$ coupling. With 2016 data, this channel can probe almost the same value as CMS \cite{Khachatryan:2015att} for $\kappa_{u\gamma}$ and can improve the limit on the  $\kappa_{c\gamma}$ by a factor of 2.
\\
\\
The direct single top quark production cross section via FCNC couplings are linearly dependent on the anomalous branching fraction as is shown in figure \ref{crossSEC}.
If the upper limits obtained in the experimental searches become better, the upper bounds on the anomalous couplings will improve significantly.
In the following we propose two variables that can be employed to find better bounds on the new physics signal cross section. 
\begin{figure}[t]
\centering 
\includegraphics[width=.48\textwidth]{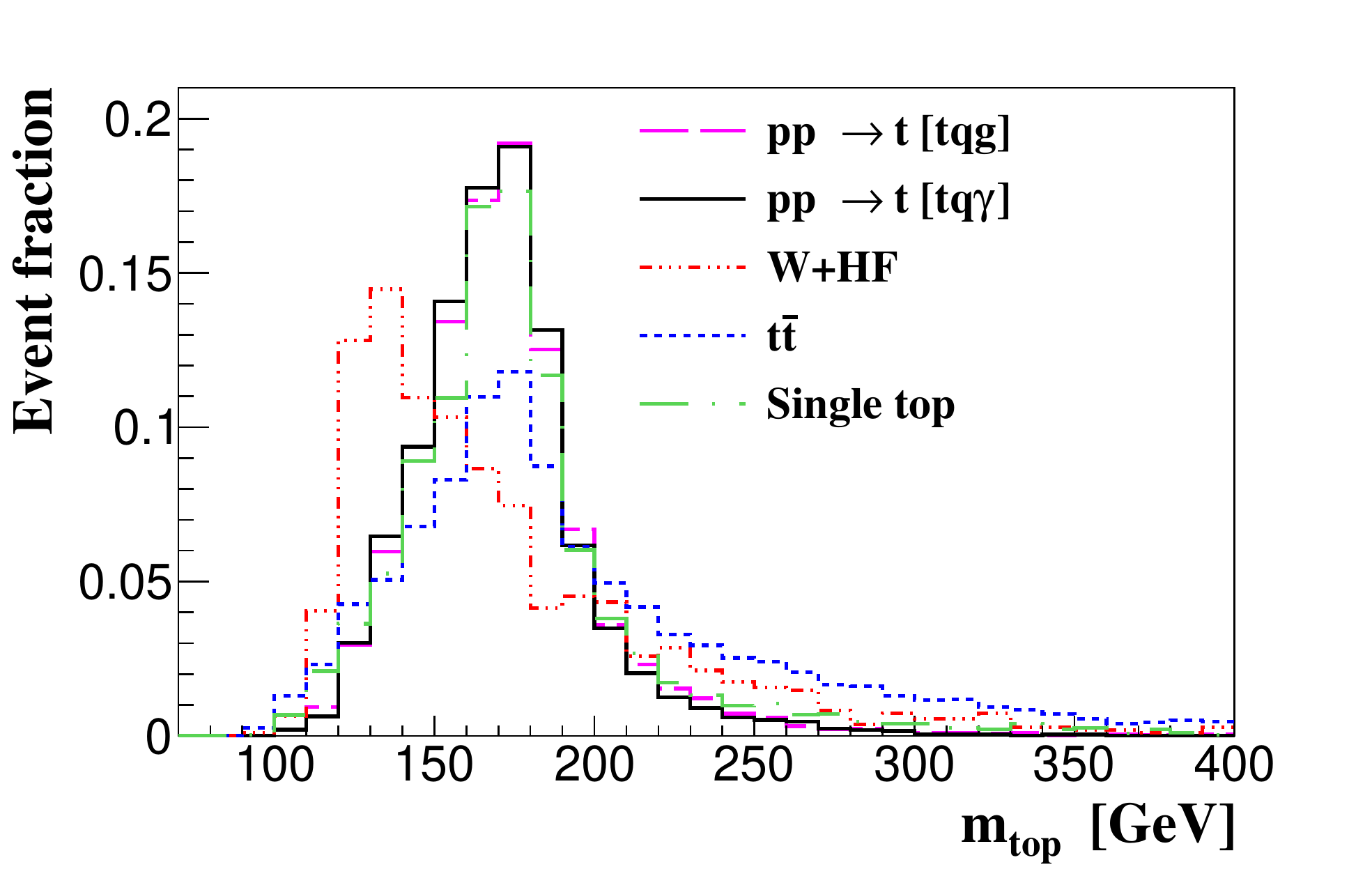}
\includegraphics[width=.48\textwidth]{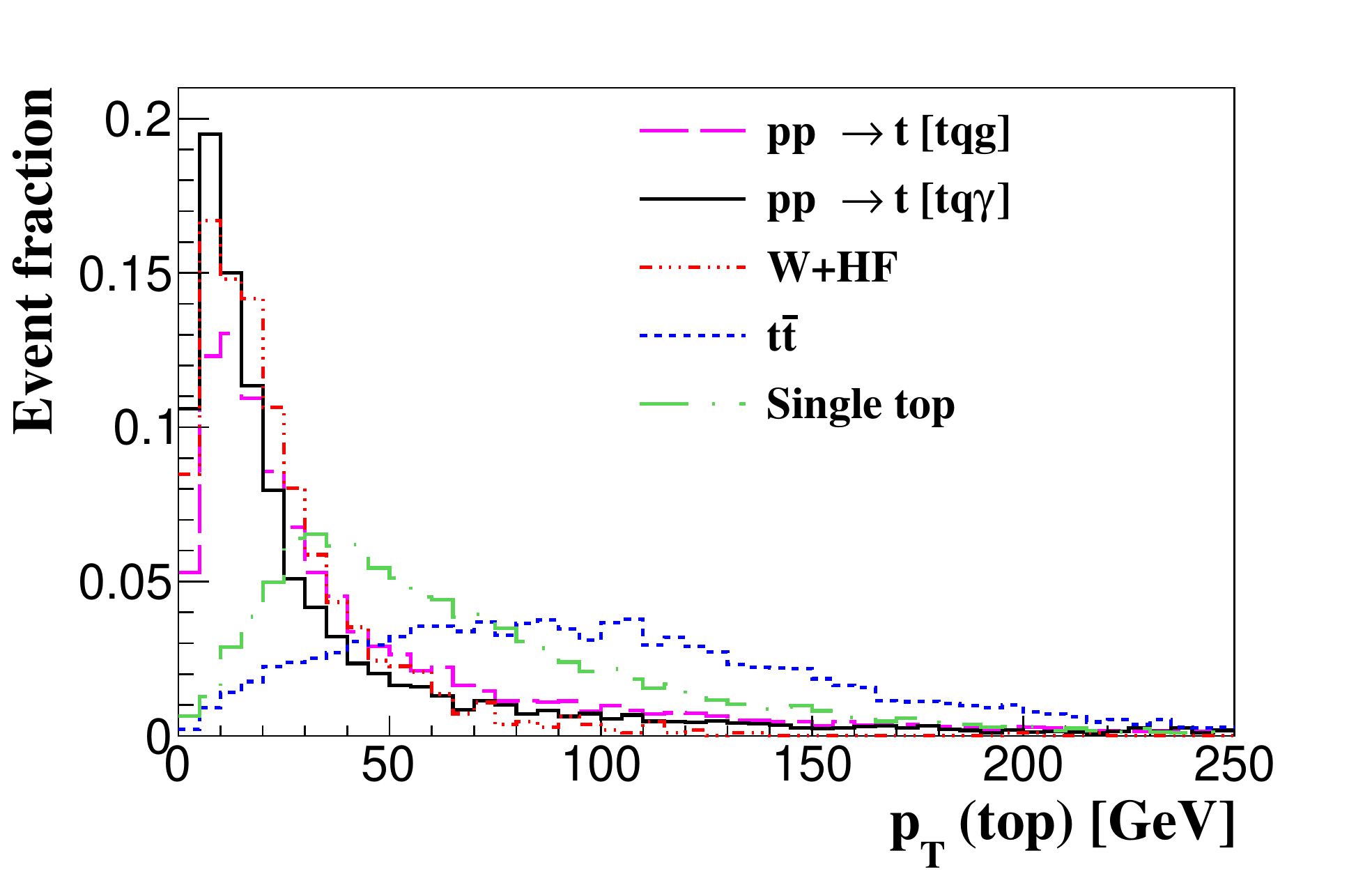}
\hfill
\caption{Distributions of the reconstructed top quark mass (left) and top quark transverse momentum (right) for signal processes (pp $\rightarrow$ t via tq$\gamma$ and tqg FCNC interactions), W+HF, single top and t$\bar{\text{t}}$ production. }
\label{topvars}
\end{figure}  
\\
\\
The contribution of the various SM backgrounds in the signal region are reported by the ATLAS collaboration in \cite{Aad:2015gea}. The  production of the W boson in association with heavy flavor quarks (W + HF) constitutes around 60\% of the whole SM background followed by the single top and t$\bar{\text{t}}$ with  13\% and 11\%, respectively.
For W + HF production, a 50\% uncertainty on the normalization is found which affects the upper limit significantly. Therefore, rejecting the W + HF  events increases the search sensitivity in the signal region.
One of the most important discriminating variable between W+HF and direct single top quark production is the top quark mass as shown in figure  \ref{topvars} (left).
Although the transverse mass of the top quark is used in the NN training in ref. \cite{Aad:2015gea} the top quark mass presents a better discrimination power. This variable can also be used in the signal region selection criteria. By selecting events with reconstructed top mass greater than 140 GeV, ~35\% of the W + HF  events are rejected while ~10\% of signal events are lost. 
In addition to the top quark mass, the transverse momentum of the top quark is a good discriminant variable between the signal and t$\bar{\text{t}}$ events.
The top quarks produced through SM processes have harder $p_T$ spectrum compared to direct top production through the FCNC interactions as shown in figure  \ref{topvars} (right). This variable can also be used in the neural network training.

\section{Conclusions}
In this study we introduced the direct single top quark production as a powerful tool to probe the tq$\gamma$ FCNC interactions. 
The most important feature of this proposal is that experimental searches can use the single top quark final state without requiring the presence of the scattered proton in the forward region spectrometers as was proposed by previous studies \cite{Ovyn:2008gs,Sun:2014qoa,Koksal:2013fta}. Therefore, the search can be done in the presence of the high number of pileups and high luminosity as it is the case at the LHC, and can benefit from the large number of events which are being recorded by ATLAS and CMS experiments at 13 TeV.  
It was shown that the direct single top quark production via tqg and tq$\gamma$  FCNC interactions leads to the same final states with similar event kinematics for final state particles. Therefore, one experimental search can be used to probe these two top quark FCNC interactions simultaneously.
\\
\\
In addition to the direct single top quark production, the pp $\rightarrow$ t+jet process is also examined for the tq$\gamma$ FCNC search with the same final state.
Although considering pp $\rightarrow$ t+jet process as a complementary signal channel improved the sensitivity of the search, one can benefit further from this channel if one selects final states with top quark and extra jet.
The CMS collaboration has done an analysis for probing the  tqg FCNC interactions through the pp $\rightarrow$ t+jet channel \cite{CMS:2014ffa} which can be used to constrain the tq$\gamma$ FCNC couplings with a similar analysis as we have done here for the pp $\rightarrow$ t process. 
\\
\\
The results of the search for direct single top production through tqg FCNC interactions performed by ATLAS collaboration are reinterpreted to set constrain on the tq$\gamma$ FCNC anomalous couplings.
Using 20.3 fb$^{-1}$ of data collected by ATLAS experiment at 8 TeV, the following limits at 95\% CL are obtained on the FCNC branching fractions:  ${\cal B}$(t $\rightarrow$ u$\gamma) < 0.07\%$ and ${\cal B}$(t $\rightarrow$ c$\gamma) < 0.19\%$. It was shown that the limits will reach ${\cal B}$(t $\rightarrow$ u$\gamma) < 0.05\%$ and ${\cal B}$(t $\rightarrow$ c$\gamma) < 0.14\%$ if all signal channels related to the tq$\gamma$ FCNC interactions are combined.
Furthermore, the sensitivity of the direct single top quark process for excluding the anomalous  tq$\gamma$ couplings are predicted at $\sqrt{s}=$ 13 TeV  for different luminosity scenarios. 
Finally, two powerful discriminant variables which are not used in the ATLAS analysis for the neural network training are introduced and discussed.

\clearpage
\providecommand{\href}[2]{#2}


\begin{thebibliography}{99}

\bibitem{ATLAS:2014wva} 
  ATLAS, CDF, CMS and D0 Collaborations,
  \textit{``First combination of Tevatron and LHC measurements of the top-quark mass''},
   [\href{http://arXiv.org/abs/1403.4427}{{\tt arXiv:1403.4427}}].

\bibitem{Glashow:1970gm} 
  S.~L.~Glashow, J.~Iliopoulos and L.~Maiani,
  \textit{``Weak Interactions with Lepton-Hadron Symmetry''},
  Phys.\ Rev.\ D {\bf 2}, 1285 (1970).

\bibitem{AguilarSaavedra:2004wm} 
  J.~A.~Aguilar-Saavedra,
  \textit{``Top flavor-changing neutral interactions: Theoretical expectations and experimental detection''},
  Acta Phys.\ Polon.\ B {\bf 35}, 2695 (2004)
  [\href{http://arxiv.org/abs/hep-ph/0409342}{{\tt arXiv:hep-ph/0409342}}].
  
\bibitem{Goldouzian:2014nha} 
  R.~Goldouzian,
  \textit{``Search for top quark flavor changing neutral currents in same-sign top quark production''},
  Phys.\ Rev.\ D {\bf 91}, no. 1, 014022 (2015)
  [\href{http://arxiv.org/abs/1408.0493}{{\tt arXiv:1408.0493}}].  
  
\bibitem{Khatibi:2015aal} 
  S.~Khatibi and M.~M.~Najafabadi,
  \textit{``Top quark flavor changing via photon''},
    Nucl.\ Phys.\ B {\bf 909}, 607 (2016)
  [\href{http://arxiv.org/abs/1511.00220}{{\tt arXiv:1511.00220}}].  
  
\bibitem{Achard:2002vv} 
   L3 Collaboration,
  \textit{``Search for single top production at LEP''},
  Phys.\ Lett.\ B {\bf 549}, 290 (2002)
  [\href{http://arxiv.org/abs/hep-ex/0210041}{{\tt arXiv:hep-ex/0210041}}].  
  
\bibitem{Abramowicz:2011tv} 
  ZEUS Collaboration,
  \textit{``Search for single-top production in $ep$ collisions at HERA''},
  Phys.\ Lett.\ B {\bf 708}, 27 (2012)
  [\href{http://arxiv.org/abs/1111.3901}{{\tt arXiv:1111.3901}}].  

\bibitem{Abe:1997fz} 
  CDF Collaboration,
  \textit{``Search for flavor-changing neutral current decays of the top quark in $p \bar{p}$ collisions at $\sqrt{s} = 1.8$ TeV''},
  Phys.\ Rev.\ Lett.\  {\bf 80}, 2525 (1998).
  
\bibitem{Abazov:2011qf} 
  D0 Collaboration,
  \textit{``Search for flavor changing neutral currents in decays of top quarks''},
  Phys.\ Lett.\ B {\bf 701}, 313 (2011)
  [\href{http://arxiv.org/abs/1103.4574}{{\tt arXiv:1103.4574}}].  

\bibitem{Aaltonen:2008qr} 
  CDF Collaboration,
  \textit{``Search for top-quark production via flavor-changing neutral currents in W+1 jet events at CDF''},
  Phys.\ Rev.\ Lett.\  {\bf 102}, 151801 (2009)
  [\href{http://arxiv.org/abs/0812.3400}{{\tt arXiv:0812.3400}}].  

\bibitem{Abazov:2010qk} 
  D0 Collaboration,
  \textit{``Search for flavor changing neutral currents via quark-gluon couplings in single top quark production using 2.3 fb$^{-1}$ of $p\bar{p}$ collisions''}
  Phys.\ Lett.\ B {\bf 693}, 81 (2010)
  [\href{http://arxiv.org/abs/1006.3575}{{\tt arXiv:1006.3575}}].  

\bibitem{Aad:2012tfa} 
  ATLAS Collaboration,
  \textit{``Observation of a new particle in the search for the Standard Model Higgs boson with the ATLAS detector at the LHC''},
  Phys.\ Lett.\ B {\bf 716}, 1 (2012)
    [\href{http://arxiv.org/abs/1207.7214}{{\tt arXiv:1207.7214}}].  
  
\bibitem{Chatrchyan:2012xdj} 
  CMS Collaboration,
  \textit{``Observation of a new boson at a mass of 125 GeV with the CMS experiment at the LHC''},
  Phys.\ Lett.\ B {\bf 716}, 30 (2012)
  [\href{http://arxiv.org/abs/1207.7235}{{\tt arXiv:1207.7235}}].  
  
    
\bibitem{Goldouzian:2014xfa} 
  R.~Goldouzian [for the CDF, D0, ATLAS and CMS Collaborations],
  \textit{``Search for FCNC in top quark production and decays''},
  7th International Workshop on Top Quark Physics (Top 2014)
  [\href{http://arxiv.org/abs/1412.2524}{{\tt arXiv:1412.2524}}]. 

\bibitem{Aad:2015gea} 
  ATLAS Collaboration,
  \textit{``Search for single top-quark production via flavour-changing neutral currents at 8 TeV with the ATLAS detector''},
  Eur.\ Phys.\ J.\ C {\bf 76}, no. 2, 55 (2016)
    [\href{http://arxiv.org/abs/1509.00294}{{\tt arXiv:1509.00294}}]. 

\bibitem{Khachatryan:2015att} 
  CMS Collaboration,
  \textit{``Search for anomalous single top quark production in association with a photon in pp collisions at $ \sqrt{s}=8 $ TeV''},
  JHEP {\bf 1604}, 035 (2016)
  [\href{http://arxiv.org/abs/1511.03951}{{\tt arXiv:1511.03951}}]. 

\bibitem{Chatrchyan:2013nwa} 
  CMS Collaboration,
  \textit{``Search for Flavor-Changing Neutral Currents in Top-Quark Decays $t \to Zq$ in $pp$ Collisions at $\sqrt{s}=8$ TeV''},
  Phys.\ Rev.\ Lett.\  {\bf 112}, no. 17, 171802 (2014)
  [\href{http://arxiv.org/abs/1312.4194}{{\tt arXiv:1312.4194}}]. 
  
\bibitem{CMS:2015xqa} 
  CMS Collaboration,
  \textit{``Search for top quark decays t $\rightarrow$ qH with H $\rightarrow$ $\gamma\gamma$ in pp collisions at $\sqrt{s}$ = 8 TeV''},
  CMS-PAS-TOP-14-019.

\bibitem{Aad:2015pja} 
  ATLAS Collaboration,
  \textit{``Search for flavour-changing neutral current top quark decays $t\to Hq$ in $pp$ collisions at $\sqrt{s}=8$ TeV with the ATLAS detector''},
  JHEP {\bf 1512}, 061 (2015)
  [\href{http://arxiv.org/abs/1509.06047}{{\tt arXiv:1509.06047}}]. 

\bibitem{deFavereaudeJeneret:2009db} 
  J.~de Favereau de Jeneret {\it et al.},
  \textit{``High energy photon interactions at the LHC''},
  [\href{http://arxiv.org/abs/0908.2020}{{\tt arXiv:0908.2020}}]. 
  
\bibitem{Fayazbakhsh:2015xba} 
  S.~Fayazbakhsh, S.~T.~Monfared and M.~M.~Najafabadi,
  \textit{``Top Quark Anomalous Electromagnetic Couplings in Photon-Photon Scattering at the LHC''},
  Phys.\ Rev.\ D {\bf 92}, no. 1, 014006 (2015),
  [\href{http://arxiv.org/abs/1504.06695}{{\tt arXiv:1504.06695}}].   

\bibitem{deFavereaudeJeneret:2008hf} 
  J.~de Favereau de Jeneret and S.~Ovyn,
  \textit{``Single top quark photoproduction at the LHC''},
  Nucl.\ Phys.\ Proc.\ Suppl.\  {\bf 179-180}, 277 (2008)
  [\href{http://arxiv.org/abs/0806.4886}{{\tt arXiv:0806.4886}}].   


\bibitem{Ovyn:2008gs} 
  S.~Ovyn and J.~de Favereau de Jeneret,
  \textit{``High energy single top photoproduction at the LHC''},
  Nuovo Cim.\ B {\bf 123}, 1126 (2008)
  [\href{http://arxiv.org/abs/0806.4841}{{\tt arXiv:0806.4841}}]. 

\bibitem{Sun:2014qoa} 
  H.~Sun,
  \textit{``Probe anomalous tqγ couplings through single top photoproduction at the LHC''},
  Nucl.\ Phys.\ B {\bf 886}, 691 (2014)
  [\href{http://arxiv.org/abs/1402.1817}{{\tt arXiv:1402.1817}}]. 
    

\bibitem{Koksal:2013fta} 
  M.~Köksal and S.~C.~Inan,
  \textit{``Anomalous $tq\gamma$ couplings in $\gamma p$ collision at the LHC''},
  Advances in High Energy Physics, Volume 2014, Article ID 935840,
  [\href{http://arxiv.org/abs/1305.7096}{{\tt arXiv:1305.7096}}].   
  
\bibitem{Inan:2014mua} 
  S.~C.~Inan,
  \textit{``Dimension-six anomalous $tq\gamma$ couplings in $\gamma\gamma$ collision at the LHC''},
  Nucl.\ Phys.\ B {\bf 897}, 289 (2015)
  [\href{http://arxiv.org/abs/1410.3609}{{\tt arXiv:1410.3609}}].     
  


\bibitem{Budnev:1974de} 
  V.~M.~Budnev, I.~F.~Ginzburg, G.~V.~Meledin and V.~G.~Serbo,
  \textit{``The Two photon particle production mechanism. Physical problems. Applications. Equivalent photon approximation''},
  Phys.\ Rept.\  {\bf 15}, 181 (1975).

\bibitem{Martin:2004dh} 
  A.~D.~Martin, R.~G.~Roberts, W.~J.~Stirling and R.~S.~Thorne,
  \textit{``Parton distributions incorporating QED contributions''},
  Eur.\ Phys.\ J.\ C {\bf 39}, 155 (2005)
  [\href{http://arxiv.org/pdf/hep-ph/0411040.pdf}{{\tt arXiv:hep-ph/0411040}}].       

\bibitem{Schmidt:2015zda} 
  C.~Schmidt, J.~Pumplin, D.~Stump and C.-P.~Yuan,
  \textit{``CT14QED PDFs from Isolated Photon Production in Deep Inelastic Scattering''},
  [\href{http://arxiv.org/abs/1509.02905}{{\tt arXiv:1509.02905}}].         
  
\bibitem{Ball:2013hta} 
  NNPDF Collaboration,
  \textit{``Parton distributions with QED corrections''},
  Nucl.\ Phys.\ B {\bf 877}, 290 (2013)
  [\href{http://arxiv.org/abs/1308.0598}{{\tt arXiv:1308.0598}}].           
  
\bibitem{Alloul:2013bka} 
  A.~Alloul, {\it et al.},
  \textit{``FeynRules  2.0 - A complete toolbox for tree-level phenomenology''},
  Comput.\ Phys.\ Commun.\  {\bf 185}, 2250 (2014)
  [\href{http://arxiv.org/abs/1310.1921}{{\tt arXiv:1310.1921}}].           

\bibitem{Degrande:2011ua} 
  C.~Degrande, {\it et al.},
  \textit{``UFO - The Universal FeynRules Output''},
  Comput.\ Phys.\ Commun.\  {\bf 183}, 1201 (2012)
  [\href{http://arxiv.org/abs/1108.2040}{{\tt arXiv:1108.2040}}].           

\bibitem{Alwall:2014hca} 
  J.~Alwall, {\it et al.},
  \textit{``The automated computation of tree-level and next-to-leading order differential cross sections, and their matching to parton shower simulations''},
  JHEP {\bf 1407}, 079 (2014)
  [\href{http://arxiv.org/abs/1405.0301}{{\tt arXiv:1405.0301}}].           
  
\bibitem{Buckley:2014ana} 
  A.~Buckley, {\it et al.},
  \textit{``LHAPDF6: parton density access in the LHC precision era''},
  Eur.\ Phys.\ J.\ C {\bf 75}, 132 (2015)
    [\href{http://arxiv.org/abs/1412.7420}{{\tt arXiv:1412.7420}}].           
  
\bibitem{CMS:2014ffa} 
  CMS Collaboration,
  \textit{``Search for anomalous Wtb couplings and top FCNC in t-channel single-top-quark events''},
  CMS-PAS-TOP-14-007.
  
\bibitem{Czakon:2013goa} 
  M.~Czakon, P.~Fiedler and A.~Mitov,
  \textit{``Total Top-Quark Pair-Production Cross Section at Hadron Colliders Through $O(α\frac{4}{S})$,''},
  Phys.\ Rev.\ Lett.\  {\bf 110}, 252004 (2013)
   [\href{http://arxiv.org/abs/1303.6254}{{\tt arXiv:1303.6254}}].
  
\bibitem{Cacciari:2008gp} 
  M.~Cacciari, G.~P.~Salam and G.~Soyez,
  \textit{``The Anti-k(t) jet clustering algorithm''},
  JHEP {\bf 0804}, 063 (2008)
  [\href{http://arxiv.org/abs/0802.1189}{{\tt arXiv:0802.1189}}].           

\bibitem{Sjostrand:2014zea} 
  T.~Sjöstrand {\it et al.},
  \textit{``An Introduction to PYTHIA 8.2''},
  Comput.\ Phys.\ Commun.\  {\bf 191}, 159 (2015)
  [\href{http://arxiv.org/abs/1410.3012}{{\tt arXiv:1410.3012}}].           

\bibitem{delphes}
 J. de Favereau, {\it et al.} ,  \textit{DELPHES 3: a modular framework for fast simulation of a generic collider experiment }, \emph{JHEP} {\bf 02} (2014) 057 [\href{http://arxiv.org/abs/1307.6346}{{\tt arXiv:1307.6346}}].
  
\bibitem{Feindt:2006pm} 
  M.~Feindt and U.~Kerzel,
   \textit{``The NeuroBayes neural network package''},
  Nucl.\ Instrum.\ Meth.\ A {\bf 559}, 190 (2006).


  
\end{thebibliography}
\end{document}